\documentclass[wrr,draft]{agutex}

\usepackage{url}
\usepackage{graphicx} 
\usepackage{lineno}

\authorrunninghead{K{\"O}HLI ET AL.}

\titlerunninghead{FOOTPRINT OF SOIL MOISTURE SENSING WITH COSMIC-RAY NEUTRONS}

\authoraddr{Corresponding author: Martin Schr{\"o}n,
(martin.schroen@ufz.de)}

\begin{document}

\title{Footprint Characteristics Revised for Field-Scale Soil Moisture Monitoring with Cosmic-Ray Neutrons}

\authors{M. K{\"o}hli, \altaffilmark{1,3}
M. Schr{\"o}n, \altaffilmark{2,3}
M. Zreda,\altaffilmark{4}
U. Schmidt,\altaffilmark{1}
P. Dietrich,\altaffilmark{2}
S. Zacharias\altaffilmark{2}
}

\altaffiltext{1}{Physikalisches Institut, Heidelberg University,}
\altaffiltext{2}{Department of Monitoring and Exploration Technologies, UFZ - Helmholtz Centre of Environmental Research, Leipzig,}
\altaffiltext{3}{These authors contributed equally to this work,}
\altaffiltext{4}{Department of Hydrology and Water Resources, University of Arizona.}


\begin{abstract}
Cosmic-ray neutron probes are widely used to monitor
environmental water content near the surface.
The method averages over tens of hectares and is unrivaled in serving representative data for
agriculture and hydrological models at
the hectometer scale.
Recent experiments, however, indicate that the sensor response to 
environmental heterogeneity 
is not fully understood.
Knowledge of the support volume 
is a prerequisite for the proper
interpretation and validation of hydrogeophysical data.
In a previous study, several physical simplifications
have been introduced into a neutron transport model
in order to derive the characteristics of the cosmic-ray probe's footprint.
We utilize a refined source and energy spectrum for cosmic-ray neutrons
and simulate their response to a variety of environmental conditions.
Results indicate that the method is particularly sensitive to 
soil moisture in the first tens of meters around the probe,
whereas the radial weights are changing dynamically with ambient water. 
The footprint radius ranges from 130 to $240\,$m 
depending on air humidity, soil moisture and vegetation.
The moisture-dependent penetration depth of 15 to 83\,cm 
decreases exponentially with distance to the sensor.
However, the footprint circle remains almost isotropic in complex terrain
with nearby rivers, roads or hill slopes.
Our findings suggest that a dynamically weighted average of point measurements
is essential for accurate calibration and validation.
The new insights will have important impact on signal interpretation,
sensor installation, data interpolation from mobile surveys,
and the choice of appropriate resolutions
for data assimilation into hydrological models.\\
An edited version of this paper was published by AGU. Copyright 2015 American Geophysical Union.\\ 
M. K\"ohli, M. Schr\"on, et al., (2015), Water Resources Research, 51, 5772-5790, DOI 10.1002/2015WR017169.
\end{abstract}

\begin{article}


\section{Introduction}

Whenever hydrology, agriculture, or climate science are concerned,
the endeavour to find efficient methods of quantifying water resources
is vitally important.
Extensive monitoring of soil moisture and above-ground  
water storage is of key importance to 
constrain hydrological model predictions or to control 
management systems for irrigation. 
However, small-scale variability of
soil moisture has always been an issue for 
the interpretation and application of point measurements
\citep{Vereecken2008,Biswas2014}. 
At large scales, remote sensing methods provide near-surface estimates of
soil-moisture. However, drawbacks are shallow penetration depth,
low temporal resolution and significant influence of surface conditions
\citep[e.g.][]{Wagner2007}.
From the modeler's perspective, 
information at scales other than the 
modeling scale requires procedures for rescaling
which introduce uncertainty during the assimilation process
\citep{Vereecken2007}. 

The method of cosmic-ray neutron sensing (CRNS) \citep{Zreda2008,Zreda2012}
has proven to be effective in serving representative data at relevant scales.
The reported footprint radius of $\approx300\,$m \citep{DesiletsZreda2013}
is much larger than spatial correlation lengths
of soil moisture patterns, typically ranging between 30 and 60\,m \citep{Western2004}.
Thereby, this technology outperforms conventional in-situ measurements
in terms of representativeness for scales beyond several tens of meters.

Neutron radiation is omnipresent in the atmosphere 
as it is generated by a nearly constant incoming flux
of cosmic rays. The presence of hydrogen
near or in the ground reduces the neutron abundance in a predictable way.
Especially the density of fast neutrons in air can serve as an efficient proxy for the
quantity of ambient water.
The continuous monitoring of background radiation 
is a passive and non-invasive solution 
to the problem of representativeness, 
because the integral average over local water sources 
is an intrinsic property of the method.

Generally, spatial integration in the support volume of a measurement
is intrinsic to most instruments in hydrogeophysics. 
When it comes to interpretation and validation, however,
accurate knowledge of the spatial extent and sensitivity to environmental
conditions is indispensable.
For that reason, investigating an instrument's support volume
is an active field of research in hydrogeophysics,
e.g. considering time-domain reflectometry \citep{Ferre1996, Ferre1998},
ground-penetrating radar \citep[see][for a review]{Huisman2003} and
nuclear magnetic resonance \citep{Legchenko2002, LubczynskiRoy2004} among others.
Methods based on the global positioning system \citep{Larson2008}
or gravimetry \citep{Creutzfeldt2010, KazamaOkubo2009}
exhibit footprints comparable to the cosmic-ray neutron probe.
However, the exact
spatial sensitivity often remains unclear and thus limits the interpretation
of measurements.
In planetary space science, investigating an instrument's footprint
is of fundamental importance, for instance, to improve the resolution and interpretation
of gamma or neutron measurements \citep{Lawrence2003, Maurice2004}.
Monte Carlo simulations were consulted to inquire the
geophysical support volume literally in depth \citep{McKinney2006}.

Using neutron transport modeling based on the Monte Carlo method, footprint characteristics of the CRNS
technique were presented initially by \citet{Zreda2008}
and investigated in detail by \citet{DesiletsZreda2013} for idealized environmental conditions.
The latter laid an important foundation to plan and improve sampling strategies
and local site arrangement. According to \citet{Zreda2012},
coastal transect experiments
confirmed the reported footprint radius of 
several hundreds of meters, but the
detailed interpretation of these measurements appears to be challenging. 
Recent investigations with mobile neutron detectors suggest that the sensor
responds to remote water bodies below the accepted theoretical distance.
Furthermore, during the course of preliminary investigations the authors could observe an effect of
extraordinary sensitivity to the very first meters around the sensor.
Thus, doubts about the accepted exponential decrease are raised
by measuring 
close to a shoreline or a small group of people,
or comparing signals of many co-located sensors in a small patch.
By using an alternative neutron source in the simulation, \citet{Rosolem2013}
found that the detector is sensitive to water vapor in heights above the probe
ranging from 412\,m to 265\,m for dry and wet air, respectively.
Their results 
indicate that the assumptions on the modeled neutron source are decisive.

A more detailed understanding of the sensor's support volume becomes important as research
projects expand to complicated terrain and mobile applications \citep[e.g.][]{Dong2014}.
An increasing number of CRNS probes are covering heterogeneous land 
which is often partly equipped with soil moisture monitoring networks
\citep[e.g.][]{Han2014, Hawdon2014, Zhu2015}.
Previous studies focused on the applicability and evaluation
of the CRNS method, where spatio-temporal conditions have mostly been
homogenous \citep[e.g.][]{Franz2012}.
Thereby it was difficult to identify invalid assumptions
on the spatial sensitivity when point measurements were averaged.
\citet{Bogena2013} identified this problem and applied a horizontally 
weighted average based on simulations from \citet{Zreda2008}, but the authors did not compare it to the non-weighted average and so the open question remains whether this approach is advantageous.
\citet{Coopersmith2014} introduced 
Voronoi-weights to differently vegetated parts of the footprint. 
However, the distance to the cosmic-ray neutron sensor was not accounted for.
In all these cases, a proper spatial weighting concept based on distance, depth 
and environmental conditions may lead to improved matching between
the cosmic-ray derived soil moisture and the averaged validation data.
Moreover, \citet{Franz2013} show that large heterogeneous structures 
in the footprint can affect the average soil moisture signal apparent to
sensors, because the neutron density and water content are non-linearly related
\citep{Desilets2010}. This phenomenon again indicates that a proper
spatial weighting of dry and wet spots could help to
compensate for heterogeneity in the field.

To address the controversy about the footprint mentioned above as well as 
the needs for an accurate weighting function, we aim to minimize
the number of physical simplifications in the numerical model. 
For example, many previous studies restrict simulations of neutron transport
to very dry conditions in air or soil \citep[e.g.][]{Zreda2012, Franz2013, Zweck2013}
and thus neglect the enormous influence of even small hydrogen sources
to the fate of neutrons.
Moreover, we are paying particular attention to a proper choice of the neutron source as model input.

In the past years various types of neutron source models have been chosen
for particle transport simulations in order to study local effects of cosmic-ray neutron interactions.
A common approach is to mimic incoming galactic cosmic rays by locating
a neutron source at $\approx8$\,km altitude
and by sampling the neutron energies randomly from a primary cosmic-ray spectrum
in a regime much above $100\,$MeV \citep[e.g.][]{Franz2013, Rosolem2013, Zweck2013}.
On the other hand, \citet{DesiletsZreda2013} and \citet{Shuttleworth2013} applied an artificially
distributed source below ground and sample from an evaporation
spectrum that peaks at $\approx1\,$MeV. This approach makes the
consequential assumption of a suppositional incoming cosmic-ray spectrum
that is comprised of high-energy neutrons only.
We will discuss both strategies in section \ref{questspectrum}.
The short review shows that there is no clear agreement about proper
energies and the location of the source. But as \citet{GlasstoneEdlund1952} and
\citet{DesiletsZreda2013} argue, neutron transport is
highly sensitive to the neutron's initial energy.

\begin{figure}[t]
\centering
\noindent\includegraphics[width=\linewidth]{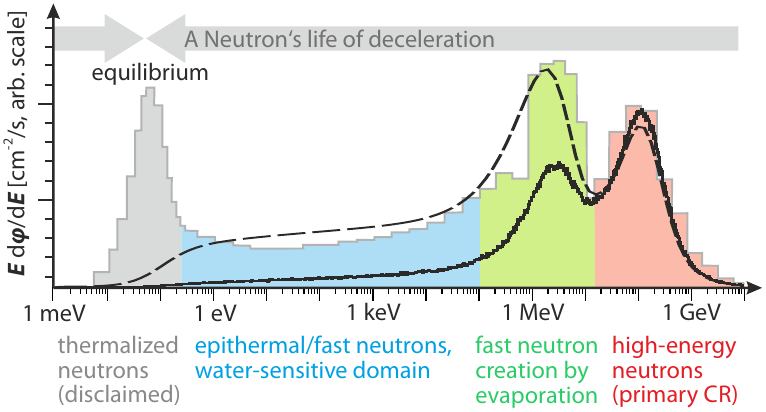}
\caption{Neutron energy spectra at the surface:
exemplary measurement by \citet{Goldhagen2002} (grey)
and simulated by \citet{sato2006} (dashed).
After subtracting the ground reflected component over pure water,
we obtain a pure incoming component (continuous black line)
which is used as the source spectrum in this study.
Colors illustrate the deceleration of initial 
high-energy neutrons (red) which interact with heavy atoms leading to the 
evaporation spectrum (green). Energy is lost by elastic collisions
with light atoms in the regime where the detector is 
particularly sensitive (blue)
until neutrons arrive energetically in a thermal equilibrium (light grey).
}
\label{goldhagen2004spectrum}
\end{figure}

Elaborated studies about atmospheric particle transport 
led to important progress in finding a
reliable energy spectrum for cosmic-ray neutrons.
\citet{sato2006} and \citet{Sato2008} simulated cosmic rays in the 
atmosphere covering a wide range of altitudes, cutoff-rigidities 
(roughly correlated to latitudes) and solar modulation potential.
Analytical descriptions for neutron energy spectra are provided 
that have been validated with independent measurements.
By choosing a parametrized energy spectrum of this kind, \citet{Lifton2014}
recently resolved some long lasting discrepancies among scaling models
for cosmogenic nuclide production.

In the same manner, we utilize the full available
energy spectrum from \citet{sato2006} near the ground to refine
previous neutron transport calculations. The objective of this study is to
specify the footprint volume and radial sensitivity for various environmental conditions from arid to humid climate.
We further investigate some of the open questions regarding the influence of 
topography, terrain, and water content in soil, air and vegetation. 


\section{Theory}
\label{theory}

Recognizing the complexity of environmental neutron physics
is essential to interpreting observations and simulations.
The neutron's sensitivity to specific types of atoms can be high, but it further 
depends on neutron energy which in turn decreases with every interaction
\citep{Rinard1991}.
For instance, even small changes in the abundance of hydrogen
can dramatically alter neutron interactions.
Model simplifications, while necessary, 
should be introduced only with great care
and their possible effects need to be assessed.

\subsection{Primary Cosmic Rays}

Cosmic radiation that is pounding the Earth originates
mostly in our galaxy, e.g. from acceleration in shock regions
of supernova remnants (see \citet{Blasi2014} for a review).
Protons are the main part of the particle flux,
accompanied by other charged nuclei. The energy spectrum of the
\emph{primary cosmic rays} peaks at around 1\,GeV per nucleon.
Depending on their momentum, cosmic-ray particles may pass the
geomagnetic fields of the Sun and the Earth. The solar magnetosphere
leads to temporal variations of the cosmic-ray intensity based on
the solar activity index. The planetary magnetosphere 
prevents cosmic rays from entering the atmosphere by
deflecting particles below an energy-dependent cutoff rigidity $r_c$.
Both effects further decrease the typical energies
of incoming radiation to several hundreds of MeV per nucleon
\citep[e.g.][]{Nesterenok2013, Grimani2011}.

\subsection{Neutron Generation in the Atmosphere}
\label{neugenatm}

\emph{Secondary cosmic-ray} particles (e.g., muons, protons, pions, neutrons)
are generated by electromagnetic and nuclear interactions mostly 
in the outer part of the Earth's atmosphere. Their intensity peaks
at the Pfotzer maximum (50--100\,g/cm$^2$ atmospheric depth,
\citet{Pfotzer1936}) and decreases exponentially by several orders
of magnitude towards sea level. However, altitudinal effects on the shape of the
energy spectrum appear to be marginal
\citep[e.g.][]{Nesterenok2013, sato2006, Hands2009, Kowatari2005, Lei2005}.
Typically, high-energy protons induce spallation of
nitrogen or oxygen nuclei in the atmosphere \citep[e.g.][]{Letaw1991}. 
This reaction releases a couple of neutrons
which are in turn able to trigger further cascades. The physics of high-energy neutron interactions
is not well-known and thus the attempt to describe the complete process is accompanied by uncertainty.

\subsection{Energy Reduction by Air, Soil and Water}
\label{energyreduction}

Above thermal energies, neutrons lose energy with every collision and cannot 
accelerate to higher energies due to their neutral electrical charge.
The final energy spectrum for neutrons at ground level is depicted in
Fig. \ref{goldhagen2004spectrum}, where three peaks are prominent.
Highly energetic neutrons at $\approx100$\,MeV (red)
are produced by intra-nuclear cascades and pre-equilibrium processes \citep{CEM1983}. 
When high-energy neutrons or protons interact with air or soil,
the excited nuclei evaporate (i.e. release) the so-called  \emph{fast} neutrons at a lower energy.
This process manifests itself at the peak at $\approx1$\,MeV (green)
and shows additional absorption fine structure due to distinct resonances of non-hydrogen
atoms.
Neutron interactions in the sub-MeV region (blue) are dominated by elastic collision,
in which, as a rule of thumb, the energy loss is correlated to the mass of the target nucleus.
Due to the extraordinarily low mass of hydrogen, this energy band
is most sensitive to water and organic molecules and thus most relevant
for the CRNS method.
Below $\approx1\, $eV the target is usually
in thermal equilibrium with the environment.
Here, the target's energy significantly contributes to the neutron's energy when transfered during a collision. As a consequence,
neutrons finally become thermalized at $k_\mathrm{B}T \approx 25\, $meV (grey),
where $k_\mathrm{B}$ is the Boltzmann constant.
Since neutrons cannot leave the thermal equilibrium towards 
relevant energies, this work disclaims the lower part of the spectrum.

\begin{figure}[t]
\centering
\noindent\includegraphics[width=\linewidth]{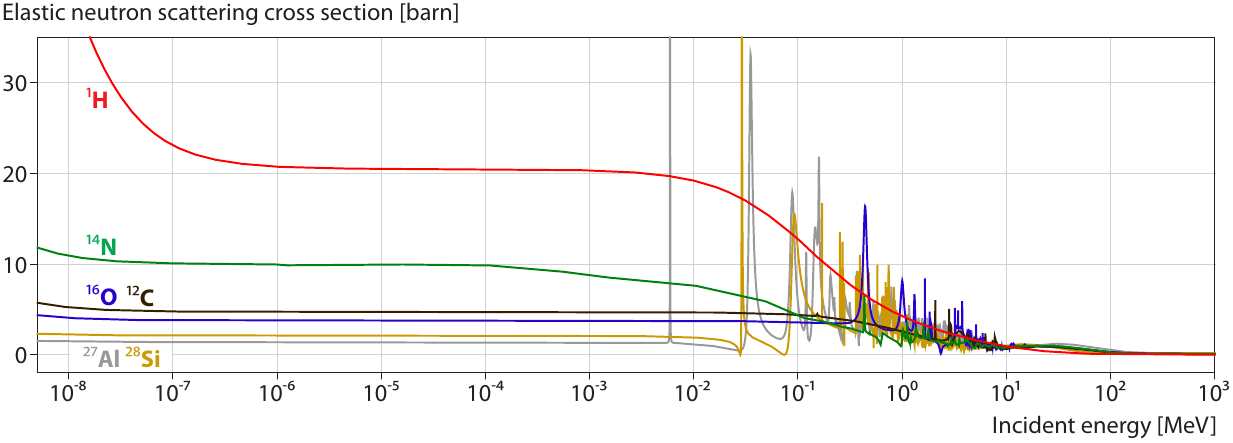}
\caption{Comparison of elastic neutron cross sections of hydrogen (red),
nitrogen (green), oxygen (blue), carbon (black), silicon (ocher),
and aluminum (grey) for kinetic energies between 5\,meV and 1000\,MeV, 
data taken from JENDL/HE-2007 \citep{jendlRef}.}
\label{crosssections}
\end{figure}

The probability of a neutron to interact with an atom
is quantified by the term \emph{cross section}.
It exhibits large variations due to isotopic composition and atomic number.
The atomic mass $A$ classifies types of possible interactions.
Very heavy nuclei with $A>80$ like $^{235}$U,
can undergo fission and scatter fast neutrons inelastically.
Intermediate nuclei of $25<A<80$ are able to absorb or scatter
the neutron inelastically. Light nuclei with $A<25$
predominantly perform elastic scattering of neutrons.
The hydrogen nucleus, $A=1$, exhibits exceptionally 
high absorption and elastic scattering cross sections. 
However, absorption is only significant in the thermal energy regime to which the ideal detector for measuring soil moisture is insensitive.
Figure \ref{crosssections} shows neutron elastic scattering cross
sections for elements that are most abundant in terrestrial air, water and solids.
Water vapor, oxygen and nitrogen
are particularly responsible for the neutron's deceleration in air.
Therefore the range a neutron can travel before thermalization
is expected to increase with altitude (i.e. decreasing air density) and
decrease with increasing air humidity. On the other hand, dense soils, organic matter
or soil water content are expected to reduce the penetration depth.

\subsection{Spatial Mixing and its Analytical Description}
\label{analyticaltassumptions}

Evaporated neutrons perform a random walk type of propagation,
because the angular probability distribution of their interactions is mostly isotropic.
In contrast, incoming high-energy neutrons do preferably interact in a forward directed process and thus vanish into the soil
without reflection.
In air, fast neutrons typically travel several tens of meters
between collisions. And since hydrogen is most efficient
in absorbing energy from the neutron, whereas air is less, information about
remote water bodies can quickly propagate within hectometers.
This process leads to 
a nearly homogeneous neutron density
in the air which can be sampled locally by the cosmic-ray neutron sensor
and represents an average of ambient hydrogen abundance in soil, air and vegetation.

As a first order approach, one could expect neutrons to behave as
a diffusive gas, as it was formulated by
\citet{GlasstoneEdlund1952}, and 
applied to a footprint estimate by \citet{DesiletsZreda2013}
besides the modeling. 
But since every collision with a particle results in an energy loss
for the neutrons, their mean free path between collisions changes and
diffusion theory loses validity. The \emph{Fermi age theory} 
\citep[e.g. applied in][]{Barkov1957}
accounts for these energy losses in a diffusive system, but 
analytical solutions exist only for mono-energetic particles and
are not feasible for the cosmic-ray neutron spectrum exposed to
a wide range of environmental atoms with different cross sections.

For these reasons, the assumption of homogeneous diffusion as applied by
\citet{DesiletsZreda2013} will bias results toward neutrons with higher energies.
Furthermore, the single-layer diffusion approach neglects the influence of the soil.
Since neutrons interact with the soil and its water content on their
path to the sensor, their energy is reduced more efficiently
compared to the propagation in air. Therefore, the homogenous 
analytical approach overestimates the horizontal footprint radius and
is rather valid for a vertical footprint above the surface.
The phenomenon of
exceptionally high vertical footprints was 
shown experimentally
with a detector on a helicopter by M. Zreda (not published) 
and indirectly with numerical simulations \citep{Rosolem2013}.

It is not feasible to join the complex problem of neutron
transport, multi-energetic Fermi age theory and two-layer 
diffusion theory into a deterministic solution.
Therefore statistical and numerical approaches are the only way
to include all necessary factors involved.

\subsection{In Quest of a Proper Model Input}
\label{questspectrum}

Simulations of cosmic-ray neutrons near the ground require consolidated
knowledge of the incoming radiation. However,
the location of the source is commonly traded against computational effort,
whereas the initial energy spectrum is bonded to a variety of uncertainties.

A popular approach is to launch secondary cosmic-ray neutrons at
$\approx8$\,km altitude and to perform their propagation through the atmosphere
\citep[e.g.][]{Franz2013, Rosolem2013, Zweck2013}.
This strategy and related simplifications come with several drawbacks:
\begin{enumerate}
\item Cross-sections of high-energy neutrons exhibit uncertainties of up 
	to 50\,\% depending on element and type of reaction, though there has been progress 
	in the last two decades \citep[e.g.][]{salvatores1994, salvatores2007}. 
	As a consequence, inconsistencies are apparent throughout
	different codes for galactic and atmospheric cosmic ray transport
	\citep[e.g.][]{Lin2012, Lei2005, sheu2003}.
	
\item Measurements of cosmic-ray energy spectra are additionally 
	accompanied by observational uncertainties.
	Comparative studies of Monte Carlo codes show differences of
	up to 20\,\% for calculating sensitivities of the neutron response
	to experimental devices \citep{Barros2014}
	and as well for the spectrum unfolding technique \citep{Ruehm2014}. 
\item The exclusive neutron source at the top of the modeled atmosphere 
	inadvertently neglects neutron generation throughout the atmosphere by
	other secondary particles like protons, pions and muons.
\item Atmospheric water vapor is often ignored, although hydrogen
	is the main moderator for neutrons.
\item The large difference in scale of the domain requires high
	computational effort to reach sufficient statistics.
\end{enumerate}
Models which rely on particle propagation through the upper atmosphere
incorporate a high complexity and vulnerability to such uncertainties involved. 

In the attempt to reduce computational effort,
other studies identified the high-energy component of the cosmic-ray
neutron spectrum as the precursor for the generation of 
fast neutrons in the soil \citep{Zreda2008, DesiletsZreda2013, Shuttleworth2013}.
Since the attenuation process of high-energy neutrons in the ground is known,
it seems likely that an artificial source in the soil
is sufficient to mimic the evaporative production of relevant neutrons.
However, some drawbacks of this method are important to note:
\begin{enumerate}
\item Attenuation of high-energy neutrons in the soil follows
	an exponential decrease that is dependent on soil type and location on Earth
	\citep{GossePhillips2001}.
\item There is no verified energy spectrum for neutrons in the soil.
	Evaporation neutrons are a significant part, but do not make up the
	spectrum as a whole (see Fig. \ref{goldhagen2004spectrum}).
\item In reality, the incoming energy spectrum from the atmosphere exhibits low-energy components and
	particularly neutrons which already evaporated in the air. 
	On entering the soil, at least one interaction is needed to alter the direction
	back to the surface.
    In contrast, the artificially generated neutrons
	in the soil can escape without any interaction.
\end{enumerate}

Considering only the evaporative neutrons in the soil can be a decent approach,
especially for dry conditions. 
However, this strategy tends to
overestimate average neutron energies, 
as incident low-energy neutrons from the top are neglected, 
and thus also overestimates the footprint size.
Moreover, the deduced footprint appears to be insensitive to soil moisture,
because its influence on neutron moderation is underestimated.

In this work, a different approach is applied, which aims to 
combine the advantages as well as avoid the drawbacks
of both strategies mentioned above.
To minimize the uncertainties of the propagated energy spectrum,
this study focuses on the domain close to the surface
by using validated results from independent atmospheric simulations
as model input.
This concept is computationally efficient and represents an
established approach in planetary space science \citep[e.g.][]{Tate2013}.

Cosmic ray propagation in the atmosphere has been
 modeled thoroughly by \citet{sato2006}. They provide a reliable 
energy spectrum of cosmic-ray neutrons for a variety of altitudes,
cutoff-rigidities, solar modulation potential and surface conditions.
These simulations have been validated with various independent measurements
at different altitudes and locations on Earth. Moreover,
the analytical formulations of the spectra
turned out to be effective in use for subsequent calculations.
The presented energy-dependent flux $\phi(E)$
is described by a mean basic spectrum $\phi_\mathrm{B}$,
a function for neutrons below 15\,MeV $\phi_\mathrm{L}$, an extension 
for thermal neutrons (disclaimed in this work), 
and a modifier $f_\mathrm{G}$ for the geometry of the interface:
\begin{linenomath*}
\[
	\phi\left(s, r_c, d, E,w \right) = 
	\phi_\mathrm{B}\left(s, r_c, d, E \right)
	\cdot f_\mathrm{G}(E,w) 
	\cdot\phi_\mathrm{L}\left(s, r_c, d \right)\,,
\]
\end{linenomath*}

where $w$ is the weight fraction of water in the ground. We focus our study on parameters for the atmospheric depth near sea level, $d = 1020$\,g/cm$^2$,
solar maximum conditions $s = 1700$\,MV and an exemplary cutoff rigidity of $r_c = 10$\,GV.
This selection might introduce small differences for different
places on Earth. However, \citet{Goldhagen02004altlat} show
that geomagnetic latitude has only very small effects on the shape of the spectrum. 
It depends slightly on atmospheric depth,
as discussed by \citet{sato2006} and
found by various authors \citep[e.g.][]{sheu2003}. However,
this is only significant for altitudes above several kilometers 
(see also section \ref{neugenatm}). 

As such spectra generally consist of an incoming
as well as a backscattered component,
the appropriate incident spectrum was separated as follows.
Firstly, for the given spectrum the response spectrum is calculated over pure water ($w = 1$), where the incoming component is dominant and thus the  uncertainties of the calculation are minimal.
Tracing the neutrons allows to determine an energy dependent multiplicity function $m(E)$ which allows to separate incoming ($m=1$) from scattered parts ($m>1$) of the spectrum. This filter can now be used to "subtract" only backscattered neutrons from the original spectrum.
A thus recalculated spectrum contains only incident neutrons 
and can be used as the source of incoming radiation for any surface condition. It is provided in the supplementaries of this manuscript.

As an exception to the otherwise isotropic distribution,
emission angles of high-energy neutrons above $10$\,MeV
are highly collimated along the downward facing direction (nadir angle $\alpha$).
According to observations and 
simulations by \citet{Nesterenok2013} 
the non-uniformity of the angular spectrum $J(\alpha)$ is given by:
\begin{linenomath*}
\[
	J(\alpha) = e^{-2.4\,(1-\cos{\alpha})}\;. 
\]
\end{linenomath*}

The presented strategy combines a universal and validated input spectrum
and angular distribution for cosmic-ray neutrons with a reduced number of simplifications
and a high computational efficiency.

\begin{figure}[t]
\centering
\noindent\includegraphics[width=\linewidth]{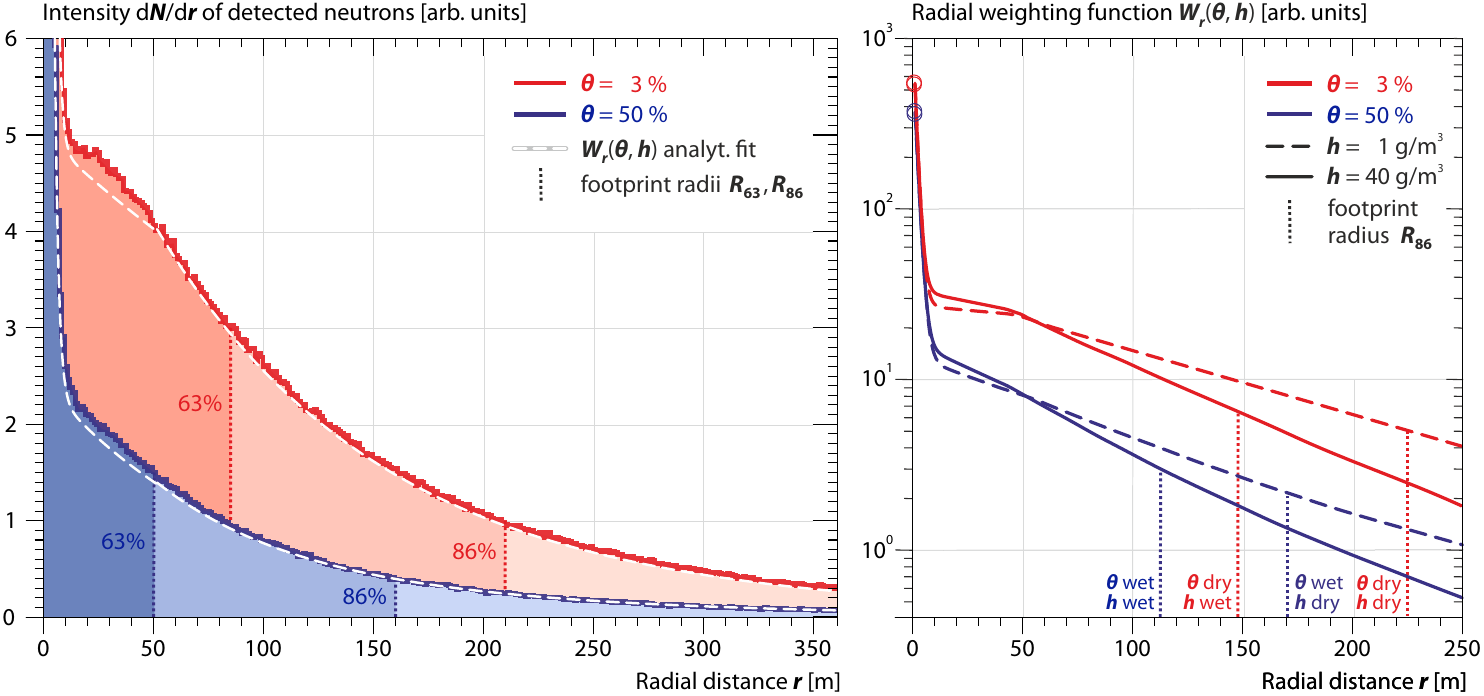}
\caption{Detected neutron intensity $\mathrm{d}N/\mathrm{d}r$ over distance $r$
between origin and detection. The analytical fit is also called \emph{radial weighting function} $W_r$. (Left):
Simulations were performed for humidity $h=10\,\mathrm{g/m}^3$ and two exemplary soil moistures
$\theta=3\,\%$ and $50\,\%$. 
Quantiles declare footprint radii $R_{63}$ and $R_{86}$ (dotted)
for $1-e^{-1}\approx63\,\%$ and $1-e^{-2}\approx86\,\%$ cumulative counts, respectively.
Peaks at $r<10$\,m reach 57.5 for $\theta=3\,\%$ and 37.9 for $\theta=50\,\%$ according to the chosen scale. (Right): Comparison of $W_r$ and the corresponding $R_{86}$ for four extreme cases of soil moisture and humidity. Both quantities are shaping the curves differently.}
\label{intenrangequant}
\end{figure}

	\subsection{Footprint Definition}

The footprint of a geophysical instrument generally covers the
area in which the medium of interest is probed and the carrier of such
information is detected. The scenario of a centrally located 
sensor which detects neutrons isotropically exhibits 
point symmetry and thus leads to the assumption of a circular
footprint area, $A=\pi r^2$.
In this work we define the \emph{travel distance} $r$ as the Euclidean 
distance between the point of detection and the point
of the neutron's first contact with the ground, also denoted as \emph{origin}.
Since $r$ depends on the neutron's initial energy and number of collisions,
it can range between 0--$10^3$\,m.
Thus a quantile definition is needed to find a definite distance $R$
within which most of the detected neutrons have probed the ground. 

By assuming an exponential decay of detected neutron intensity over
travel distance, which relates to the solution of a simple diffusion model,
\citet{Zreda2008} and \citet{DesiletsZreda2013} legitimate the use of
two $e$-folding lengths, i.e. the 86\,\% quantile, in order to 
define the footprint radius. Figure \ref{intenrangequant}
illustrates the radial decrease of the detected neutron intensity $W_r$
as a result of Monte Carlo simulations performed in this work.  
Although the calculated response does not exhibit a simple exponential shape,
any other quantile would be an arbitrary choice as well.
Careful interpretation of this value is recommended, however, 
because a high quantile value will always treat long-range
neutrons with favour, regardless of how often they have probed the soil.
Nevertheless, we decide to follow the definition of the 86\,\% quantile for
historical reasons and denote the according footprint radius with $R_{86}$
and the footprint area as $A=\pi R_{86}^2$\,.

The number of neutrons $N_R$ that have originated within a distance
$R$ from the sensor is given by
\begin{linenomath*}
\begin{equation}
 N_R=\int_0^RW_r\,\mathrm{d}r\,. 
\label{n-as-integral}
\end{equation}
\end{linenomath*}
In order to find the distance within which 
86\,\% of the detected neutrons originate, the following
equation is solved for $R_{86}$ numerically:
\begin{linenomath*}
\begin{equation}
\int_0^{R_{86}}W_r\,\mathrm{d}r=0.86\int_0^\infty W_r\,\mathrm{d}r\,.
\label{obtaining-r86}
\end{equation}
\end{linenomath*}

In analogy we define the \emph{penetration depth} $D_{86}$ in the soil
as the integral of a depth weighting function $W_d$ which is
expected to also decrease with distance $r$ to the sensor.


\section{Modeling}
\label{theModeling}
Monte Carlo simulations are able to track the histories
of millions of neutrons. By taking all relevant
physical interactions into account, the summary statistics of 
a large number of neutrons can reveal insights into their collective effects.
In this study we apply the Monte Carlo method to address both large geometric scaling
and anisotropic conditions.

	\subsection{Software}
	\label{software}
To address the specific needs of neutron-only interactions, we developed 
the Ultra Rapid Adaptable Neutron-Only Simulation (URANOS)
based on the Monte Carlo approach for neutron transport.
The software was originally developed to simulate specific characteristics
of the Heidelberg neutron spin echo detectors and was
adapted to the cosmic-ray neutron problem.
The physics model follows the implementation declared by the ENDF database standard and
described by OpenMC \citep{openmcRef}, a recent Monte Carlo code alternative to MCNP. It features the treatment of elastic collisions
in the thermal, epithermal, and fast regime, as well as inelastic collisions, 
absorption and absorption-like processes (e.g. evaporation) which play a dominant role
for the given elements (these are the processes described by the ENDF MT identifiers 5, 103, 107, 208, 209, 210).
Cross sections, energy distributions and angular
distributions were taken from the databases ENDF/B-VII.1 \citep{endfRef}
and JENDL/HE-2007 \citep{jendlRef}. 

The URANOS code was tailored to the problem of
neutron transport in environmental science.
By neglecting unnecessary physical processes (e.g. fission and gamma cascades)
this leads to a significant increase in the computing speed compared to
other available Monte Carlo codes for the description of neutron transport.
In preparatory studies we explored the performance of the URANOS model
in reproducing results from standard software like MCNPX.
The tests successfully agreed 
in many different setups (not shown) such as the one presented by \citet{sato2006}.
Particular attention was turned to the reproduction of the results from
MCNPX performed by \citet{DesiletsZreda2013}. Using exactly the same setup of soil composition and source definition
we were able to reproduce the reported footprint radius of $\approx300\,$m
and confirm the negligible dependence on soil moisture.

\begin{figure}[t]
\centering
\noindent\includegraphics[width=\linewidth]{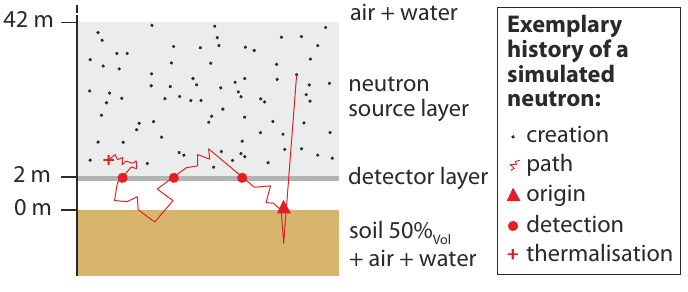}
\caption{Setup of the simulation containing a 
40\,m thick neutron source layer in the atmosphere and a thin detector layer
at 2\,m above ground. A passing neutron is counted if it had preceding contact with the soil.
The footprint is determined based on the distance between the origin and detection.}
\label{simsetup}
\end{figure}

	\subsection{Neutron Source}
	\label{theSource}
Neutrons are launched from point sources randomly distributed 
in the region from 2\,m to 42\,m above the surface (Fig. \ref{simsetup}).
Energies are sampled from a pre-calculated spectrum based on \citet{sato2006}, which is provided in the supplementaries of this manuscript.
High-energy neutrons are launched with a collimated angular distribution
(see section \ref{questspectrum}). The source intensity was chosen according to statistical errors.
More neutrons would lead to more accurate and smooth data. We experienced that $10^7$
neutrons are a reasonable trade-off between computational effort and precision.

	\subsection{Detector}
	\label{theDetector}
Neutrons are recorded individually in an horizontally infinite detector layer.
Any neutron that experienced interaction with the soil is counted as it passes the layer.
The infinite plane detector overlays the atmosphere by means of a 25\,cm
high sheet at a vertical position of 175--200\,cm. This geometry was chosen because we aim to compare our results with \citet{DesiletsZreda2013}, who tallied the neutron fluxes in a $2\,$m detector layer.
The detector layer is crossed by the neutrons and thus maps the spatial field of neutron densities. It is an appropriate abstraction of a realistic, small-scale detector volume of the same height that absorbs neutrons. As tests confirm, multiple counts of a single neutron in the detector layer account for the measured density equivalent for a single count per volume detector. This relation holds if (1) the dimension of the absorbing detector medium stays below typical scale lengths of neutron interactions (10--$100\,$m), and (2) particles do not scatter multiple times in that volume. That is very unlikely for non-thermal neutrons
and furthermore does not factorize in the count statistics.

We refer to several statements of the effective energy range to which the detector is 
sensitive. Following practical considerations by \citet{DesiletsZreda2013}
and theoretical by \citet{hertel1985}, the detection energy is set to a window from 10\,eV to $10^3$\,eV.
\citet{Kouzes2007} reports that the detection efficiency of
moderated helium-3 detectors is nearly
constant in that energy regime, which is why signal weighting for
different energies is not needed.

Recent studies reported that the common cosmic-ray neutron detectors (presented by \citet{Zreda2012})
are contaminated by $\approx30$\,\% thermal neutrons \citep{McJannet2014}.
We do not account for this issue, because this study aims to 
investigate characteristics for a detector ideally tailored to the needs
of environmental water sensing. 

\subsection{Air, Soil and Water}
\label{theDomains}

The modeled pure air medium consists of 78\,$\%_\mathrm{Vol}$ nitrogen, 21\,$\%_\mathrm{Vol}$ oxygen
and 1\,$\%_\mathrm{Vol}$ argon at a pressure of 1020\,mbar.
The soil extends to a depth of 6\,m and the air to 1000\,m.
Both, soil and air are represented by planes
of infinite extension, which can have subdomains, either to create a density profile in depth or
to add specific entities like water or a detector. 
The soil consists of 50\,$\%_\mathrm{Vol}$ solids and a scalable amount of H$_2$O.
The solid domain is comprised of 75\,$\%_\mathrm{Vol}$ SiO$_2$ and
25\,$\%_\mathrm{Vol}$ Al$_2$O$_3$ at a compound density of 2.86\,g/cm$^3$. 
Thus, the total densities vary from 1.43\,g/cm$^3$ to 1.93\,g/cm$^3$
for 0\,$\%_\mathrm{Vol}$ and 50\,$\%_\mathrm{Vol}$ soil moisture, respectively. 

Further chemical constituents regarding rock types are not significant for fast neutron moderation,
according to calculations from \citet{Zreda2012} and \citet{Franz2012a} and 
the discussion in section \ref{energyreduction}.


\section{Results and Discussion}

\begin{figure}[t]
\centering
\noindent\includegraphics[width=\linewidth]{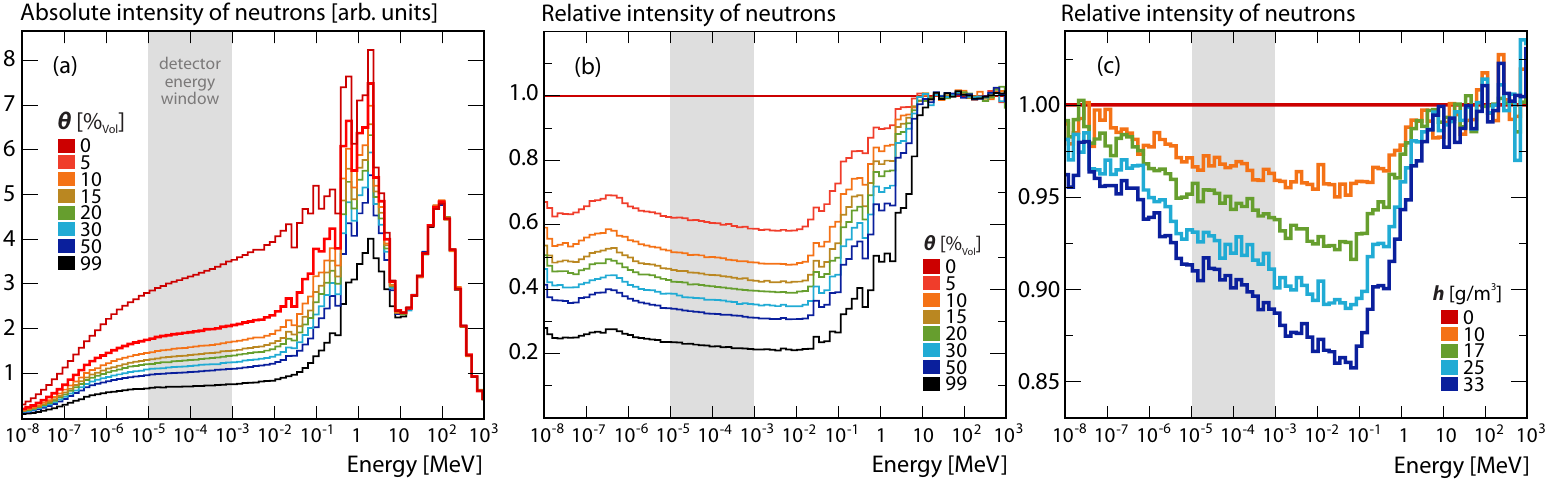}
\caption{
Calculated neutron spectra above ground with the
highlighted energy window of the detector (grey)
and the disclaimed thermal domain to its left,
(a) for different soil moistures at an air humidity of 10\,g/m$^3$,
(b) intensities of (a) scaled relative to 0\,\% volumetric soil moisture,
(c) intensities for different air humidities relative to 0\,g/m$^3$ 
at 10\,\% volumetric soil moisture.}
\label{intensitiesAboveGround}
\end{figure}

The response of the ground to the incoming
flux of cosmic-ray neutrons lead to several interesting features in
the resulting energy spectrum. Figure \ref{intensitiesAboveGround}a,b
confirms the efficient reduction of neutron intensity by soil moisture
in the relevant energy range of the CRNS method.
Figure \ref{intensitiesAboveGround}c shows
that water vapor particularly affects neutrons at the upper end of the
energies considered. In this energy domain, neutrons
cover the largest distances and are consequently
exposed to the highest path-integrated amount of air.
In general, neutrons appear to be very sensitive to small amounts 
of hydrogen in soil and air.

	\subsection{Radial Footprint Changes with Humidity and Soil Moisture}

\begin{figure}[b]
\centering
\noindent\includegraphics[width=\linewidth]{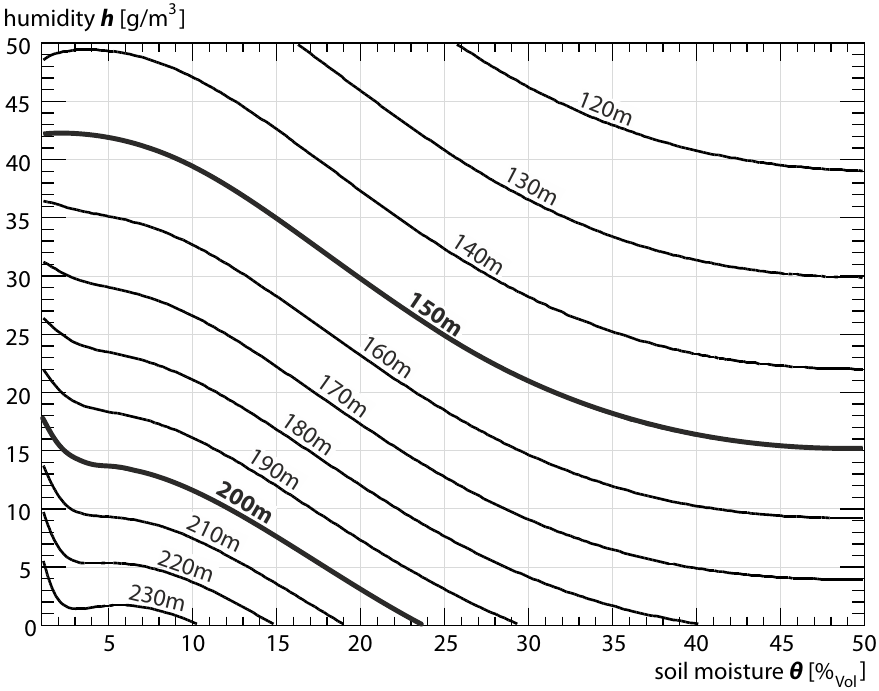}
\caption{Footprint radius $R_{86}$ (contour lines) and its dependency
on soil moisture $\theta$ 
and air humidity $h$ 
at sea level calculated by eqs. \ref{obtaining-r86} and \ref{theFormula}.
Complex response to small amounts of hydrogen is evident.
Corresponding data is provided in the supplementaries.}
\label{fp-contours}
\end{figure}

We performed simulations with a variety of volumetric water contents in the soil
from 0--50\,\% and absolute humidity in the air from
0--50$\,\mathrm{g/m}^3$. Figure \ref{intenrangequant} illustrates the sensitivity
of the detector to neutrons originating at different radial distances $r$.
This \emph{radial weighting function} $W_r$ can assist in finding
a properly weighted mean of independent soil moisture measurements. It 
further shows that little contribution is made by neutrons from $r>200\,$m
and highest contribution comes from $r<10\,$m around the sensor.

The peak at $r<10\,$m accounts for neutrons that directly emerge from the ground
and have a high probability to be detected even though most of them come
from the lower part of the neutron energy spectrum. The region up to $r<50\,$m
describes the average mean free path of most of the environmental neutrons
in humid air. 
For distances between 50--200\,m neutrons interact with the soil multiple times
until they are detected, which in turn means that with increasing $r$, average neutron
energies quickly become insufficient in order to arrive at the detector
before thermalization. From about $200\,$m on, detected neutrons are dominated
by the higher energetic part of the spectrum, which appear to be higher
in flux rates and are able to probe the soil very far from the detector. 

Due to the different neutron energies involved, 
we found an accurate fit to the intensity distribution (Fig. \ref{intenrangequant})
by splitting the radial domain into four exponential parts.
An analytical description can be obtained for $\theta\geq2\,\%$\,:
\begin{linenomath*}
\begin{equation}
W_r(h,\theta) \approx 
\left\{
  \begin{array}{lr}
    F_1\,e^{-F_2r} + F_3\,e^{-F_4r}\,, & 0.5\,\mathrm{m}< r\leq50\,\mathrm{m}\\
    F_5\,e^{-F_6r} + F_7\,e^{-F_8r}\,, & 50\,\mathrm{m}< r< 600\,\mathrm{m}\\
  \end{array}
\right.
\label{theFormula}
\end{equation}
\end{linenomath*}
where the parameter functions $F_i(h,\theta)$ are individually dependent
on humidity and soil moisture as given in appendix \ref{fp-parameters}.
The separation at $r=50\,$m accounts for the non-trivial shape
of the function as described above. For $r>50$\,m both exponential terms 
describe diffusion-like processes each accounting for soil moisture and air humidity presence.
On the contrary, in $r\leq 50$\,m diffusion is not the main process, however, since the same functional structure still holds numerically, it was chosen for convenience.
Following equation \ref{obtaining-r86} we integrate $W_r(h,\theta)$ 
numerically.
The resulting $R_{86}(h,\theta)$ is analytically difficult to grasp, thus
we illustrate the numerically integrated results as contours
in Figure \ref{fp-contours} and
present a numerical matrix in the supplementaries.
The contour plot shows that the footprint radius ranges from 
240\,m to 130\,m between arid and tropical climate, respectively. 

The response to soil moisture variations is significant for humid climate between 10--40\,$\%_\mathrm{Vol}$
as well as for very dry conditions $<3\,\%_\mathrm{Vol}$. Previous
studies underestimated the role of soil moisture for the footprint
due to the choice of a modeled neutron source below the surface
(see section \ref{questspectrum}). 
Comparative studies (not shown) indicated
that this detail is the major cause for the discrepancy
to findings from \citet{DesiletsZreda2013}.
Moreover, the decrease of the footprint with increasing
soil moisture does not necessarily 
imply that the area-average estimate is less representative.
According to \citet{korres2015}, spatial variability of soil moisture
tends to be low for rather wet soils. 
In this context, the effective representativeness of the CRNS method 
appears to be almost unchanged.

The response to variations of absolute humidity features 
a 10\,m decrease of the footprint radius for every change of 4--6\,g/m$^3$ water vapor.
\citet{Zreda2012} refers to $\approx10$\,\% reduction of the footprint from
dry to saturated air, which can easily span $\approx25$\,g/m$^3$.
This change corresponds to a 20\,\% change in footprint radius 
calculated with URANOS. However, \citet{DesiletsZreda2013}
investigated the influence of humidity in further detail and found 
a 10\,m decrease for every change of $\approx6$\,g/m$^3$ humidity from MCNPX
simulations with dry soil. This value is consistent with
results from URANOS, whereas the slightly higher gradient is a consequence
of the different energy spectra used in the models.

The function $W_r$ lays the basis for a refinement of the commonly applied sampling strategy.
The accepted method equally weights point measurements from three distinct radii
\citep{Zreda2012, Franz2012} which correspond to an exponential weighting function
(see section \ref{analyticaltassumptions}). 
In contrast, the present work shows that (1) the first tens of meters provide dominant
contribution to the signal in a rather non-exponential relation, and (2) the shape of the 
weighting function changes temporally as it is affected by variable moisture conditions.
It is therefore not possible to elaborate a universal sampling strategy. As a rule of thumb we 
recommend to take more samples closer to the probe (e.g. 0--10\,m) than was previously recommended.
Subsequently, data should be weighted with $W_r(h,\theta)$ in a post-processing mode
(see appendix \ref{appWM}).

\subsection{Uncertainty Analysis}

In the simulated system containing soil, atmosphere, and a detector, uncertainties propagate non-linearly due to the variety of neutron interactions involved. As an indication of their total effect, we analyzed uncertainties of our calculations by means of the influence on the footprint radius $R_{86}$.

Variations of cross sections by their standard deviation, given in the ENDF data base, lead to changes of $R_{86}$ by $4\,\%$, $3\,\%$, and $2\,\%$ for $\theta =3\,\%$, $10\,\%$ and $40\,\%$, respectively. The effect of elastic scattering dominates the budget by approximately $70\,\%$. Further details about this analysis are provided in the supplementaries. The errors of the cross sections can be considered as systematic for neutron transport simulations in general.
We further analyzed the impact of different source spectra as model input in a test case with $10\,\%$ soil moisture and $5\,$g/m$^3$ air humidity. As explained in section \ref{questspectrum} the incident spectrum was generated over water by subtracting the soil response from the original mixed spectrum. Variations of this soil response spectrum by $20\,\%$ alters $R_{86}$ by $2.5\,\%$. If the emission angles of source neutrons were not set according to their angular distributions, but chosen perpendicular to the surface, the change of the footprint radius would be $2.5\,\%$ applied to high energetic neutrons only and $3.0\,\%$ using sub-MeV neutrons. Compared to the uncertainties involved in our calculations the impact of other source spectrum models can be much higher. 
The integration of the counted particles (eq. \ref{obtaining-r86}) further leads to statistical uncertainties on $R_{86}$ in the order of $0.2\,\%$ for $10^7$ neutrons.

All in all we conservatively report a total error of $\Delta R_{86}=$ 4--6$\,\%$, which scales from wet to dry conditions.

	\subsection{Footprint Scaling with Vegetation and Air Pressure}
	
To investigate the footprint variability under vegetated conditions,
we modeled above-ground vegetation as a layer of height $H_\mathrm{veg}$,
containing an exemplary mixture of water and carbon with 
a density of $\rho_\mathrm{veg}=0.005$\,g/cm$^3$. This corresponds to $\approx4.4\, \mathrm{kg/m}^3$ biomass water equivalent (BWE), since we have chosen the molecular composition of the gas in a way that the living plant consists of $\approx12\,\%$ carbon by weight. For layer heights below a few meters, variations of either density or height have comparable effect on neutron moderation. Therefore, this method can provide valid estimations of the vegetation effect in terms of both, height and water equivalent.
 
\begin{figure}
\centering
\noindent\includegraphics[width=\linewidth]{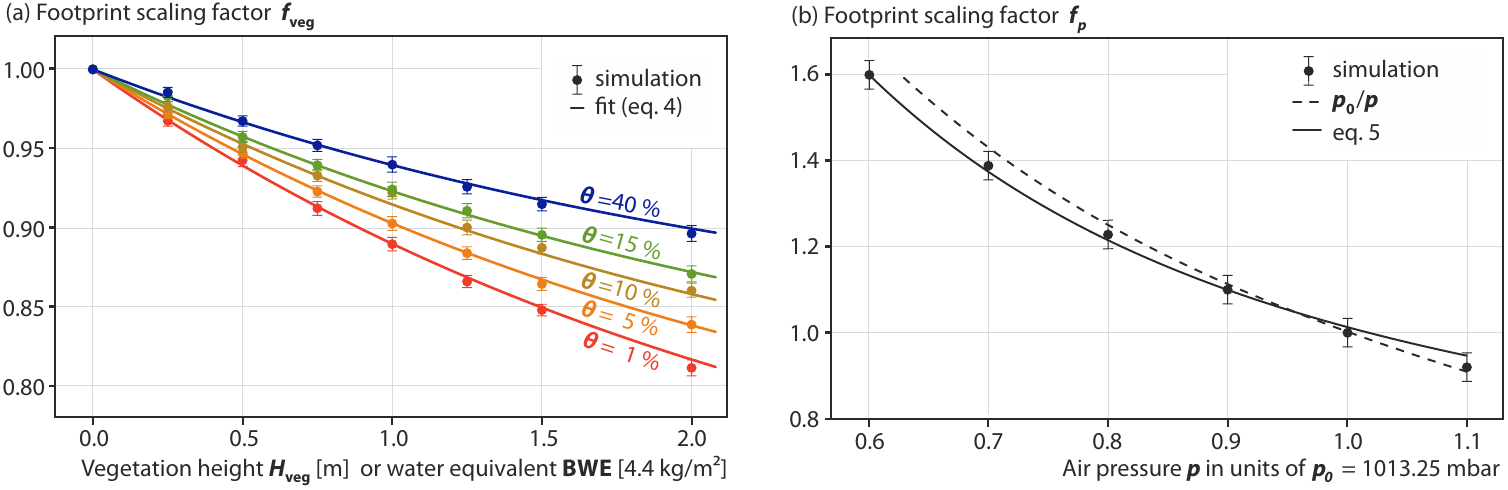}
\caption{Change of the footprint radius $R_{86}(h,\theta)$ by vegetation and air pressure. (a): The scaling factor $f_\mathrm{veg}$ is reduced by vegetation height which corresponds to a biomass water equivalent. The effect is weakened by increasing soil moisture and air humidity, the latter was fixed to $h=10\,\mathrm{g/m}^3$ in this example. (b): The scaling factor $f_p$ increases with altitude, which corresponds to decreasing air pressure. In this simulation we consider $h=5\,\mathrm{g/m}^3$ $\forall\,p$ and $\theta=5\,\%$.}
\label{scaling}
\end{figure}

From the perspective of the neutron, the layer introduces a new source of hydrogen in the air
and consequently reduces the traveling range in the same manner as humidity.
For example, the footprint radius is reduced by $\approx20$\,\% 
for crops of height $H_\mathrm{veg}=2$\,m ($\mathrm{BWE} \approx 8.8\,\mathrm{kg/m}^2$) in dry soils.
From simulations presented in Figure \ref{scaling}a we find an exponential dependence of the footprint scaling factor $f_\mathrm{veg}$ 
on vegetation height $H_\mathrm{veg}$:
\begin{linenomath*}
\begin{equation}
	f_\mathrm{veg}(\theta) = 1-0.17\left(1-e^{-0.41H_\mathrm{veg}}\right)\left(1+e^{-7\theta}\right)\,, 
\end{equation}
\end{linenomath*}
where $\theta$ is given in units of m$^3$/m$^3$. For thin vegetation cover a linearisation in $H_\mathrm{veg}$ is appropriate. As \citet{Franz2013u} demonstrate, water in above ground
biomass influences the signal in another way than 
homogeneously distributed soil moisture, which
is well reflected by the URANOS model approach (see also Fig. \ref{intensitiesAboveGround}).

On the other hand, the footprint can also expand with decreasing air pressure (e.g. increasing altitude).
The lower air density allows neutrons to cover longer distances between collisions.
For example, the footprint can be 20\,\% larger at a $\approx2000\,$m
altitude ($\simeq800\,$mbar) compared to sea level.
Although a reciprocal fit is a reasonable estimate \citep{DesiletsZreda2013},
our results presented in Figure \ref{scaling}b indicate an
exponential dependence on $p$ due to the presence of hydrogen:
\begin{linenomath*}
\begin{equation}
	f_p = \frac{0.5}{0.86-e^{-p/p_0}}\approx p_0/p\,. 
\end{equation}
\end{linenomath*}
However, differences between the two models appear to be insignificant.

By taking the scaling factors into account, the final
footprint radius can be estimated with
\begin{linenomath*}
\begin{equation}
	R_{86}(h,\theta,p,\mathrm{veg}) = f_p\cdot f_\mathrm{veg}(\theta) \cdot R_{86}(h,\theta)\,.
	\label{fp}
\end{equation}
\end{linenomath*}
An extreme case of vegetation is a forest site, where cosmic-ray
sensors are placed to study the influence of wet biomass or
interception water in the canopy \citep[e.g.][]{Desilets2010}).
In order to provide a first glimpse of the influence of a forest
on the footprint, we set up a gas representing the molecular 
composition of an exemplary forest with a density of
$\rho_\mathrm{forest}=0.0016$\,g/cm$^3$, which corresponds to $\approx1.4\, \mathrm{kg/m}^3$.
Considering $h=10\,$g/m$^3$ and $\theta=10\,\%$, our
results indicate that the sensor footprint in a forested
ecosystem is reduced to 78\,\% or 44\,\%
for canopy heights of 15\,m or 30\,m, respectively.
Qualitatively, this reduction should be taken into account when calibration or
validation of the CRNS probe is performed in forests and in different seasons
\citep[e.g.][]{Franz2013eco, Bogena2013, Lv2014}.
Future investigtions should focus on various vegetation models and cover a range of parameters in order to gain profound understanding of neutron interactions at individual agricultural or forest sites.

	\subsection{Penetration Depth in the Soil}

The thickness of the probed soil layer is an important advantage of the CRNS method
compared to most remote-sensing products. Cosmic-ray neutrons can penetrate the first decimeters of the soil 
almost unhindered, whereas electromagnetic signals
interact within the upper 0--5\,cm. \citet{Franz2012a} showed
that the effective representation of the penetration depth, $z^*(\theta)$, is a
reciprocal function of soil moisture,
but it is unclear how it varies with the distance from the probe.

In URANOS we logged the vertical positions where neutrons 
lost energy in a scattering process, i.e. probed the soil.
Above $\theta\geq10\,\%$, the penetration depth of neutrons appears to decrease
exponentially. This behaviour can be expected from a simple mono-energetic 
Beer-Lambert approach \citep{Beer1852}, and has also been found by \citet{Zreda2008}.
A simple analytical description of the vertical weighting function was found for $\theta\ge10\,\%$:
\begin{linenomath*}
\begin{equation} 
W_d(r,\theta) \propto e^{-2d/D_{86}(r,\theta)}. 
\label{Wd}
\end{equation} 
\end{linenomath*}
The relation can be used to obtain a properly averaged mean value of point
measurements when compared to the cosmic-ray derived estimates.
The numerical determination of the penetration depth $D_{86}$, however,
is certainly valid for any soil moisture condition $\theta\in(1..50\,\%)$:
\begin{linenomath*}
\begin{equation}                
D_{86}(r,\theta) = 
	\rho_\mathrm{bd}^{-1}\left( p_0+p_1\left( p_2+e^{-r/100}\right)
	\frac{p_3+\theta}{p_4+\theta}\right).
	\label{D86}
\end{equation}
\end{linenomath*}
The quantity denotes up to which depth 86\,\% of the detected neutrons
had contact with constituents of the soil. 
Numerical parameters are provided in Table \ref{parameters}, $\theta$
is in units of $\mathrm{m}^3/\mathrm{m}^3$.
\begin{figure}
\centering
\noindent\includegraphics[width=\linewidth]{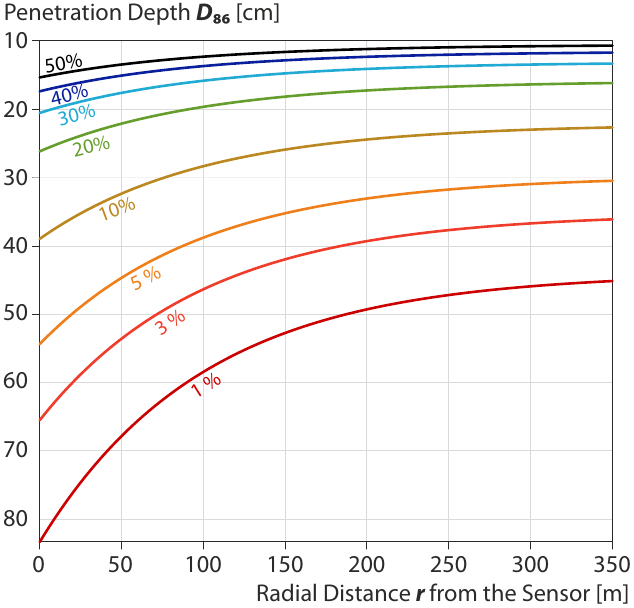}
\caption{Dependency of the penetration depth $D_{86}$ (eq. \ref{D86})
on radial distance $r$ to the sensor for a range of
soil water contents $\theta$ (coloured).
An exemplary humidity $h=10\,\mathrm{g/m}^3$ and
soil composition according to section \ref{theDomains} is considered.
}
\label{depth-contours}
\end{figure}

Figure \ref{depth-contours} shows penetration depths $D_{86}(r,\theta)$
as a function of radial distance $r$ from the sensor
for exemplary soil moisture values $\theta$.
For dry soil $D_{86}(r,\theta\approx1\,\%)$ ranges from
$83\,$cm right below the sensor to $46\,$cm
at $r=300\,$m distance. At most, the penetration depth varies between
15\,cm and 83\,cm below the sensor for wet and dry soil, respectively. This is
in close agreement with depths of 12--76\,cm given by \citet{Zreda2008}. 
The reported values are rather confirmed than contradicted by URANOS, 
because they stemmed from experiences and various studies in the research field of cosmogenic nuclide production
and are thus independent of the mentioned model approach. 
On average over the first tens of meters distance, the functional dependency on $\theta$ (eq. \ref{D86})
is relatively similar to the reciprocal model for the \emph{effective sensor depth} $z^*(\theta)$ 
from \citet{Franz2012a}. Their model was constrained on the limits introduced by \citet{Zreda2008}
and validated with measurements and hydrodynamic simulations. 
Further evidence for the correct performance of the URANOS model 
provides the comparison with measurement depths of $50-100$\,cm on the Moon or Mars missions,
where cosmic-ray neutrons penetrate dry ground of similar chemical composition \citep{Elphic2008,McKinney2006}.

	\subsection{Terrain Structures and Topography}

Cosmic-ray neutron probes are sometimes placed
close to roads, trees, rivers or in hilly terrains.
In an analogous manner mobile rover surveys inevitably
pass alongside forests, lakes or fields of different land use.
In most of these cases we do not expect an isotropic footprint of
the probe, because large structures of different hydrogen content vary throughout the viewing directions.

\begin{figure}[t]
\centering
\noindent\includegraphics[width=\linewidth]{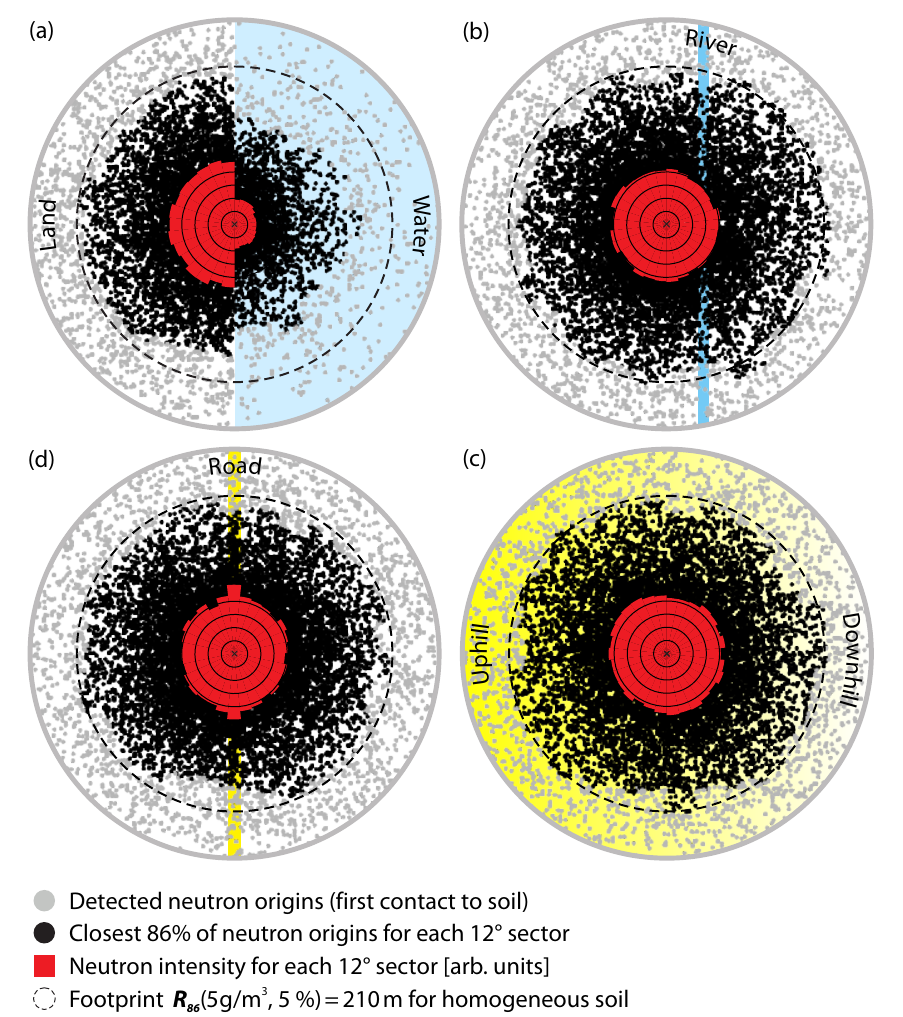}
\caption{Anisotropy of detected neutron origins (black) and 
neutron intensity (red) determined for every $12^\circ$ sector
of a circle around a centered detector. The displayed extent is 270\,m in radius,
whereas the dashed line represents the isotropic footprint with radius
$R_\mathrm{86}(h,\theta)\approx210$\,m,
considering $\theta=5\,\%$ and $h=5\,\mathrm{g/m}^3$. 
The four exemplary cases illustrate bare soil (white) with 
(a) a coast line (blue), (b) a 10\,m river at 50\,m distance,
(c) a 10\,m concrete road (yellow) and (d) a $20^\circ$ hill slope 
from the left down to the right.}
\label{anisotropy}
\end{figure}

In order to quantify the anisotropy of detected neutrons,
we simulated four exemplary cases where such scenarios 
are extreme. In Figure \ref{anisotropy} the vicinity 
of a centered detector is shown and the isotropic footprint
$R_{86}(h=5\,\mathrm{g/m}^3, \theta=5\,\%)=210\,$m is indicated (dashed line).
Dots illustrate the origin of detected neutrons, where the closest 86\,\%
of total neutrons are emphasized (black) in each direction.
We discretized the area into $12^\circ$ sectors in order to quantify
range (black dots) and intensity (red) for 30 discrete directions.

In a coast line setup (Fig. \ref{anisotropy}a) the density of
the origins (dots) and neutron intensity (red) appear to be much smaller 
in the ponded area. The range of neutrons decreases by up to 30--40\,\% although neutrons still
manage to travel long distances over water. Their contribution to
the count rate sharply drops to about 40\,\% at the interface. 

In Figure \ref{anisotropy}b the detector is placed 50\,m away from a 
10\,m wide river. This setup can be found were cosmic-ray neutron probes are located
within small catchments with creeks or irrigated land.
Neutron origins clearly show that the river
hardly contributes to the signal because most neutrons lose too much energy 
after probing water (see also point density and neutron intensity for water, Fig. \ref{anisotropy}a).
This is also visible in the intensity which shows
a slight asymmetry towards the dry side.
However, the setup reveals a slightly wider footprint 
in the direction to the river, as a consequence of the intensity gap.

A detector carried on a dry, concrete road (Fig. \ref{anisotropy}c) is
a common scenario for rover applications \citep[e.g.][]{Chrisman2013}. The sensor
detects about 10--20\,\% more neutrons per sector from the road than from other
directions. However, the decrease of the footprint along the road
due to short-range dominated contribution is marginal.
The effect of the road is expected to be weaker for tarry material,
as it contains hydrogen and carbon.

In Figure \ref{anisotropy}d we illustrate the investigation of
neutron detection under more complex topography, here being a $20^\circ$
steep hill slope.
From detailed analysis we found that the
uphill footprint (left) does not differ significantly from downhill (right),
although small asymmetries in the neutron origins occur.
Neutron intensity from uphill is about 0.26\,\% higher compared to downhill,
which is far beyond significance of the count rate.

These idealized cases demonstrate that the geometry of complex terrain only slightly influences soil moisture measurements with the cosmic-ray neutron sensor. However, the anisotropic contributions to the count rate should be investigated individually if accuracy matters. To add more reality to the scenarios, future studies on topography and structures should account for correlating quantities like gradients of air pressure, humidity, or soil moisture down the hill or close to rivers. As a consequence of more collisions and more efficient moderation, these quantities are expected to treat neutrons from uphill preferentially.

	\subsection{Experimental Evidence?}
	\label{expevidence}
	
Since the footprint definition is based on a radial symmetry,
direct empirical evidence is difficult to achieve with natural structures. However, approaching
water surfaces and transiting the coast line has been a common procedure to 
determine the range of detected neutrons. For example, \citet{Kuzhevski2003}
moved the detector over a lake and interprets that the
signal strength is hardly sensitive to neutrons from the land side at distances greater
than 200\,m. In the last years, many experiments with the COSMOS detector have been performed across a water-land boundary by the group of M. Zreda.
First data from Oceanside Pier (California, US) indicate that
the sensitive distance is on the order of 100--200\,m at sea level.

\begin{figure}[t]
\centering
\noindent\includegraphics[width=\linewidth]{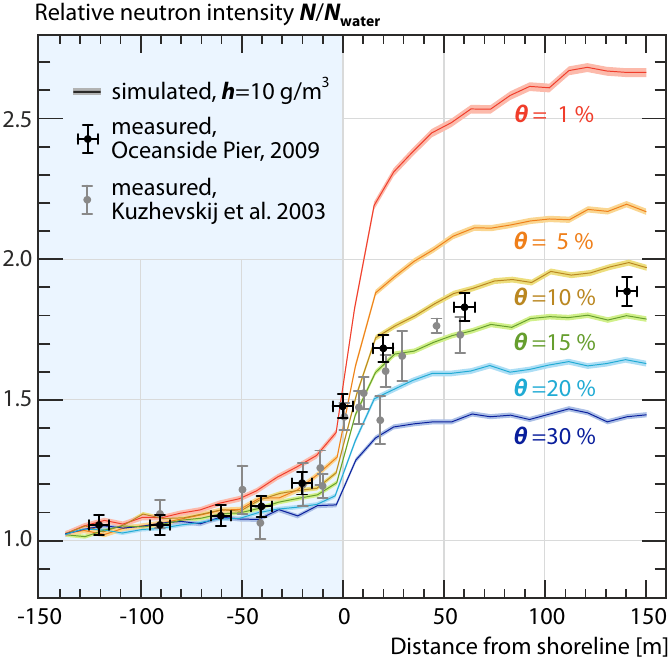}
\caption{Coastal transect experiments simulated with a 4\,m square-shaped detector
every $\pm10\,$m from the coast line. Relative neutron counts
show good agreement with measurements across a water-land boundary at the Oceanside Pier (US)
as well as tests at Lake Seliger (RU, from \citet{Kuzhevski2003}).
Air humidity $h$ and soil moisture $\theta$ were unknown.}
\label{shoreline-plot}
\end{figure}

With URANOS we made an attempt to reproduce these transect experiments
by moving a 4\,m square-shaped detector 
over pure water and land with exemplary soil moistures from 1 to 30\,\% and fixed air humidity $h=10$\,g/m$^3$.
Figure \ref{shoreline-plot} illustrates the simulations and the two experiments mentioned above.
Simulated signal strengths clearly correspond to the measurements
and give an indication of the soil water content which was unknown
at the time of the experiments.
The signal gradient is asymmetric over water (left) and land (right),
which agrees with results from \citep{Franz2013}, who 
investigated the influence of large wet structures on the signal strength.
It is further interesting to
note that $R_{86}$ ranges from 168--220\,m (according to the considered range in soil moisture, 1--30\,\%).
However, these values cannot be identified in the experiment,
because the signal is almost saturated by 150\,m. Both effects 
can be explained by
(1) the overestimation of dry over wet regions in the signal,
as a consequence of the non-linear relation: $\theta\mapsto N$ \citep{Desilets2010},
(2) the effective removal of traveling neutrons due to the presence of a
water body on their way to the detector, and (3) the non-radial geometry of the experiment.

We must conclude that transect experiments 
do not give a direct measure of the
footprint radius under conditions where the
instrument is usually applied. However, the presented data 
provide evidence for the valid performance of the URANOS model. 

\section{Summary and Conclusions}

This work investigates the footprint defined as the water-sensitive support volume of the cosmic-ray neutron sensor.
Previous simulations by \citet{DesiletsZreda2013} drew general conclusions
from a number of model assumptions and provided a decent estimate
of the footprint for dry conditions. 
The travel distance of neutrons probing the soil, however, is very sensitive to initial energies
and even to small amounts of hydrogen on their way.
As a consequence, the complexity of environmental neutron transport appears to 
impede any attempt to simplify the problem.
Therefore, we felt the need for revisiting neutron transport models and 
for addressing some of the open questions regarding the radial sensitivity, humid climate, or terrain structures. Simulations in this work 
were performed by the Monte Carlo code URANOS, whereas the concept is applicable to any standard software.
From the results of this work we draw the following conclusions:

\begin{enumerate}

\item{The revised footprint radius $R_{86}(h,\theta,p,\mathrm{veg})$ is $240\,$m (18\,ha area)
for bare soil and purely dry conditions at sea level.
However, significant influence of soil moisture $\theta$, humidity $h$ and vegetation can 
further reduce the radius by more than 40\,\%. 
In contrast, decreasing air pressure may expand it by $\approx1\,\%$ per 10\,mbar. The total error $\Delta R_{86}$ was estimated conservatively to be about 4--6$\,\%$.
The dynamic footprint has implications for methods for the interpolation of survey data,
irrigation management and data assimilation for hydrological models.}

\item{The signal strength per radial distance, $W_r(h, \theta)$, is highly non-linear in $r$, $h$ and $\theta$
and exhibits extraordinary sensitivity to the first few meters. 
As a rule of thumb, at least
half of the neutron intensity reflected by the soil is due to the first 50\,m around the sensor. 
Consequently, dynamic weighting of horizontal averages can be essential for sensor calibration
and validation with soil moisture monitoring networks.}

\item{The penetration depth $D_{86}(r,\theta)$ of detected neutrons directly below the sensor 
ranges from $\approx15$ to 83\,cm depending on soil moisture.
An exponential decrease with depth is a good estimate for the sensor's vertical sensitivity,
whereas the depth in turn shrinks significantly with radial distance to the sensor.}

\item{The circular shape of the footprint remains isotropic for most field applications,
like hilly terrain, nearby rivers or heterogeneous land.
However, large water bodies or forests nearby can reduce range and intensity of detected neutrons from that direction.
Dry roads can contribute to an overestimate of neutron counts by a few percent.
While rover surveys are often exposed to a variety of environmental conditions,
these findings can have implications for interpretation and geostatistical interpolation
of spatial data.}

\item{Transect experiments can be helpful to investigate the detector response to remote water bodies.
In the same manner they allow to validate input models and strategies for Monte Carlo driven simulations.
URANOS is able to reproduce these measurements adequately, however, this method is not appropriate to give direct evidence for the radial footprint size.}

\end{enumerate}

The present study shows that the description of the footprint and neutron intensity 
is non-trivial to an exceptional degree. For this reason it is not possible to 
conclude with an easy and straight-forward analytical formulation.
We recommend to read out values from the figures presented here or 
from numerical and highly resolved data given in the supplementaries.
Individual simulations should be performed for every probe
in order to analyze the local response specific to the surrounding environment.

Future work is needed to experimentally verify the results of this study.
For example, to test the suggested spatial sensitivity of the sensors,
the performance of weighted averages needs to be assessed
for point data from sampling campaigns or soil moisture monitoring networks.
Despite the dynamic characteristics of the footprint,
the capability to average water content over a large volume
is undisputed and remains a valuable advantage of the method of cosmic-ray neutron sensing
in the field of soil moisture monitoring.

\appendix

\section{Parameter Functions}
\label{fp-parameters}

The parameter functions $F_i(h, \theta)$ in equation \ref{theFormula} 
have been obtained empirically. Units for air humidity $h$ and
soil moisture $\theta$ are $\mathrm{g/m}^3$ and $\mathrm{m}^3/\mathrm{m}^3$, respectively.
Table \ref{parameters} contains the related numerical parameters $p_j$.
In the supplementaries we provide application-ready scripts to calculate $W_r$
in R, MatLab and Excel.
\begin{eqnarray*}
F_1 &=& p_0\left(1+p_3 h\right) e^{-p_1 \theta}+p_2\left(1+p_5 h\right)-p_4 \theta ,\\
F_2 &=& \left(\left(p_4 h-p_0\right) e^{-\frac{p_1 \theta}{1+p_5 \theta}}+p_2 \right) \left(1+p_3 h\right),\\
F_3 &=& p_0\left(1+p_3 h\right)e^{-p_1 \theta} +p_2-p_4 \theta\,, \\
F_4 &=& p_0 e^{-p_1 \theta}+p_2-p_3 \theta +p_4 h,\\
F_5 &=& p_0 \left( 0.02 - \frac{1}{p_5 ( p_6 \theta+h-p_5) }\right) \\
	&&\cdot (p_4 - \theta) e^{-p_1(\theta-p_4)} + p_2\left( 0.7 - h \theta p_3\right), \\
F_6 &=& p_0 (h + p_1) + p_2 \theta, \\
F_7 &=& \left( p_0 \left(1-p_6 h \right) e^{-p_1 \theta\left(1- p_4 h\right)}+ p_2 - p_5 \theta  \right) \cdot \left(2 + p_3 h \right), \\ 
F_8 &=& \left( (p_4 h-p_0) e^{\frac{-p_1\theta}{1 + p_5 h + p_6 \theta}}+ p_2 \right) \cdot (2 + p_3 h).
\end{eqnarray*}

\begin{table*}[t]
\label{parameters}
\caption{Parameters for $F_i$ (Appendix \ref{fp-parameters})
and $D_{86}$ (eq. \ref{D86}).}
\centering
\begin{tabular}{l | l l l l l l l}
\hline
$ $ & $p_0$ & $p_1$ & $p_2$ & $p_3$ & $p_4$ & $p_5$ & $p_6$ \\\hline
$F_1$ & $8735$ & $17.1758$ & $11720$ & $0.00978$ & $7045$ & $0.003632$ & $ $ \\
$F_2$ & $2.7925\cdot10^{-2}$ & $5.0399$ & $2.8544\cdot10^{-2}$ & $0.002455$ & $6.851\cdot10^{-5}$ & $9.2926$ \\
$F_3$ & $247970$ & $17.63$ & $374655$ & $0.00191$ & $195725$ & $ $ & $ $ \\
$F_4$ & $5.4818\cdot10^{-2}$ & $15.921$ & $0.6373$ & $5.99\cdot10^{-2}$ & $5.425\cdot10^{-4}$ & $ $ & $ $ \\
$F_5$ & $1383702$ & $4.156$ & $5325$ & $0.00238$ & $0.0156$ & $0.130$ & $1521$ \\
$F_6$ & $6.031\cdot10^{-5}$ & $98.5$ & $1.0466\cdot10^{-3}$ & $ $ & $ $ & $ $ & $ $ \\
$F_7$ & $11747$ & $41.66$ & $4521$ & $0.01998$ & $0.00604$ & $2534$ & $0.00475$ \\
$F_8$ & $1.543\cdot10^{-2}$ & $10.06$ & $1.807\cdot10^{-2}$ & $0.0011$ & $8.81\cdot10^{-5}$ & $0.0405$ & $20.24$ \\
\hline
$D_{86}$ & $8.321$ & $0.14249$ & $0.96655$ & $26.42$ & $0.0567$ & $ $ & $ $ \\
\end{tabular}
\end{table*}


\section{Weighted Mean for Soil Moisture Comparisons}
\label{appWM}

Equations \ref{theFormula} and \ref{Wd} can be used to
weight individual point measurements in order to validate
or calibrate the signal apparent to a cosmic-ray neutron sensor.

The general procedure to obtain a weighted average $\left<\theta_k\right>$
from measurements $\theta_k$ with the weighting function $W_k$ is as follows:
\begin{linenomath*}
\[ 	\left<\theta_k\right> = \frac{\sum_k \theta_k\cdot W_k}{\sum_k W_k}\,. \]
\end{linenomath*}

Let $\theta_{ij}$ be a sample of soil moisture
at the depth $d_j$ and distance $r_i$ to the cosmic-ray neutron sensor.
The field-mean soil moisture $\left<\theta_{ij}\right>$ and
humidity $\left<h\right>$ can be averaged with
the equal weight $W_{ij}=1\,\forall\,i,j$. 
We then suggest to firstly compute the vertical average $\left<\theta_j\right>_i$
of the data at each point $i$ with the weighting function
$W_{d_j}(r_i,\left<\theta_{ij}\right>)$, eq. \ref{Wd}.
Secondly, these values can be
averaged horizontally ($\forall i$) with the weighting function
$W_{r_i}(\left<h\right>,\left<\theta_{ij}\right>)$, eq. \ref{theFormula}.


\begin{acknowledgments}
Supporting animations are available online at \url{http://www.ufz.de/cosmosfootprint}. Supplementary information of this work include data for Fig. \ref{fp-contours}, scripts to calculate eq. \ref{theFormula}, measurements from M. Zreda in Fig. \ref{shoreline-plot}, the incoming neutron source spectrum, and details of the uncertainty analysis for the cross sections.
Data used to support figures \ref{goldhagen2004spectrum}, \ref{crosssections},
and \ref{shoreline-plot} can be found in the cited papers. 
URANOS was developed for the project
"Neutron Detectors for the MIEZE method"
funded by the German Federal Ministry of
Education and Research (BMBF), grant identifier: 05K10VHA.
Source code and support for URANOS can be provided by 
M. K\"ohli.
MS acknowledges kind support by the Helmholtz Impulse and
Networking Fund through Helmholtz Interdisciplinary School for
Environmental Research (HIGRADE).
The contribution of MZ has been funded through the COSMOS
project by the U.S. National Science Foundation, grant identifier: ATM-0838491.
The present work benefited from stimulating and critical
discussions with Darin Desilets (Hydroinnova LLC). MK thanks A. Nesterenok and T. Sato for fruitful discussions. MS thanks J. Mai, L. Sch\"uler, J. Weimar, and U. Wollschl\"ager for kind assistance and valuable comments.
The research was funded and supported by Terrestrial Environmental Observatories (TERENO), which is a joint collaboration program involving several Helmholtz Research Centers in Germany. 

\end{acknowledgments}

\end{article}


\begin{thebibliography}{71}
\providecommand{\natexlab}[1]{#1}
\expandafter\ifx\csname urlstyle\endcsname\relax
  \providecommand{\doi}[1]{doi:\discretionary{}{}{}#1}\else
  \providecommand{\doi}{doi:\discretionary{}{}{}\begingroup
  \urlstyle{rm}\Url}\fi

\bibitem[{\textit{Barkov et~al.}(1957)\textit{Barkov, Makarin, and
  Mukhin}}]{Barkov1957}
Barkov, L., V.~Makarin, and K.~Mukhin (1957), Measurement of the slowing down
  of neutrons in the energy range 1.46-0.025 {eV} in water, \textit{Journal of
  Nuclear Energy (1954)}, \textit{4}(1), 94 -- 102,
  \doi{http://dx.doi.org/10.1016/0891-3919(57)90124-9}.

\bibitem[{\textit{Barros et~al.}(2014)\textit{Barros, Mares, Bedogni,
  Reginatto, Esposito, F.~Goncalves, Vaz, and R\"uhm}}]{Barros2014}
Barros, S., V.~Mares, R.~Bedogni, M.~Reginatto, A.~Esposito, I.~F.~Goncalves,
  P.~Vaz, and W.~R\"uhm (2014), Comparison of unfolding codes for neutron
  spectrometry with {B}onner spheres, \textit{Radiation Protection Dosimetry},
  \textit{161}(1-4), 46--52, \doi{10.1093/rpd/nct353}.

\bibitem[{\textit{Beer}(1852)}]{Beer1852}
Beer (1852), Bestimmung der {A}bsorption des rothen {L}ichts in farbigen
  {F}l\"ussigkeiten, \textit{Annalen der Physik}, \textit{162}(5), 78--88,
  \doi{10.1002/andp.18521620505}.

\bibitem[{\textit{Biswas}(2014)}]{Biswas2014}
Biswas, A. (2014), Season- and depth-dependent time stability for
  characterising representative monitoring locations of soil water storage in a
  hummocky landscape, \textit{{CATENA}}, \textit{116}(0), 38 -- 50,
  \doi{http://dx.doi.org/10.1016/j.catena.2013.12.008}.

\bibitem[{\textit{Blasi}(2014)}]{Blasi2014}
Blasi, P. (2014), Recent results in cosmic ray physics and their
  interpretation, \textit{Brazilian Journal of Physics}, \textit{44}(5),
  426--440, \doi{10.1007/s13538-014-0223-9}.

\bibitem[{\textit{Bogena et~al.}(2013)\textit{Bogena, Huisman, Baatz,
  Hendricks~Franssen, and Vereecken}}]{Bogena2013}
Bogena, H.~R., J.~A. Huisman, R.~Baatz, H.-J. Hendricks~Franssen, and
  H.~Vereecken (2013), Accuracy of the cosmic-ray soil water content probe in
  humid forest ecosystems: The worst case scenario, \textit{Water Resources
  Research}, \textit{49}(9), 5778--5791, \doi{10.1002/wrcr.20463}.

\bibitem[{\textit{Chadwick et~al.}(2011)\textit{Chadwick, Herman, Obložinský,
  Dunn, Danon, Kahler, Smith, Pritychenko, Arbanas, Arcilla, Brewer, Brown,
  Capote, Carlson, Cho, Derrien, Guber, Hale, Hoblit, Holloway, Johnson,
  Kawano, Kiedrowski, Kim, Kunieda, Larson, Leal, Lestone, Little, McCutchan,
  MacFarlane, MacInnes, Mattoon, McKnight, Mughabghab, Nobre, Palmiotti,
  Palumbo, Pigni, Pronyaev, Sayer, Sonzogni, Summers, Talou, Thompson, Trkov,
  Vogt, van~der Marck, Wallner, White, Wiarda, and Young}}]{endfRef}
Chadwick, M., M.~Herman, P.~Obložinský, M.~Dunn, Y.~Danon, A.~Kahler,
  D.~Smith, B.~Pritychenko, G.~Arbanas, R.~Arcilla, R.~Brewer, D.~Brown,
  R.~Capote, A.~Carlson, Y.~Cho, H.~Derrien, K.~Guber, G.~Hale, S.~Hoblit,
  S.~Holloway, T.~Johnson, T.~Kawano, B.~Kiedrowski, H.~Kim, S.~Kunieda,
  N.~Larson, L.~Leal, J.~Lestone, R.~Little, E.~McCutchan, R.~MacFarlane,
  M.~MacInnes, C.~Mattoon, R.~McKnight, S.~Mughabghab, G.~Nobre, G.~Palmiotti,
  A.~Palumbo, M.~Pigni, V.~Pronyaev, R.~Sayer, A.~Sonzogni, N.~Summers,
  P.~Talou, I.~Thompson, A.~Trkov, R.~Vogt, S.~van~der Marck, A.~Wallner,
  M.~White, D.~Wiarda, and P.~Young (2011), {ENDF/B-VII.1} nuclear data for
  science and technology: Cross sections, covariances, fission product yields
  and decay data, \textit{Nuclear Data Sheets}, \textit{112}(12), 2887 -- 2996,
  \doi{http://dx.doi.org/10.1016/j.nds.2011.11.002}, special Issue on
  ENDF/B-VII.1 Library.

\bibitem[{\textit{Chrisman and Zreda}(2013)}]{Chrisman2013}
Chrisman, B., and M.~Zreda (2013), Quantifying mesoscale soil moisture with the
  cosmic-ray rover, \textit{Hydrology and Earth System Sciences},
  \textit{17}(12), 5097--5108, \doi{10.5194/hess-17-5097-2013}.

\bibitem[{\textit{Coopersmith et~al.}(2014)\textit{Coopersmith, Cosh, and
  Daughtry}}]{Coopersmith2014}
Coopersmith, E.~J., M.~H. Cosh, and C.~S. Daughtry (2014), Field-scale moisture
  estimates using {COSMOS} sensors: A validation study with temporary networks
  and leaf-area-indices, \textit{Journal of Hydrology}, \textit{519, Part
  A}(0), 637 -- 643, \doi{http://dx.doi.org/10.1016/j.jhydrol.2014.07.060}.

\bibitem[{\textit{Creutzfeldt et~al.}(2010)\textit{Creutzfeldt, G\"untner,
  Vorogushyn, and Merz}}]{Creutzfeldt2010}
Creutzfeldt, B., A.~G\"untner, S.~Vorogushyn, and B.~Merz (2010), The benefits
  of gravimeter observations for modelling water storage changes at the field
  scale, \textit{Hydrology and Earth System Sciences}, \textit{14}(9),
  1715--1730, \doi{10.5194/hess-14-1715-2010}.

\bibitem[{\textit{Desilets and Zreda}(2013)}]{DesiletsZreda2013}
Desilets, D., and M.~Zreda (2013), Footprint diameter for a cosmic-ray soil
  moisture probe: Theory and {M}onte {C}arlo simulations, \textit{Water
  Resources Research}, \textit{49}(6), 3566--3575, \doi{10.1002/wrcr.20187}.

\bibitem[{\textit{Desilets et~al.}(2010)\textit{Desilets, Zreda, and
  Ferr\'e}}]{Desilets2010}
Desilets, D., M.~Zreda, and T.~Ferr\'e (2010), Nature's neutron probe: Land
  surface hydrology at an elusive scale with cosmic rays, \textit{Water
  Resources Research}, \textit{46}(11), \doi{10.1029/2009WR008726}.

\bibitem[{\textit{Dong et~al.}(2014)\textit{Dong, Ochsner, Zreda, Cosh, and
  Zou}}]{Dong2014}
Dong, J., T.~E. Ochsner, M.~Zreda, M.~H. Cosh, and C.~B. Zou (2014),
  Calibration and validation of the cosmos rover for surface soil moisture
  measurement, \textit{Vadose Zone Journal}, \textit{13}(4), --,
  \doi{doi:10.2136/vzj2013.08.0148}.

\bibitem[{\textit{Elphic et~al.}(2008)\textit{Elphic, Chu, Hahn, James,
  Lawrence, Prettyman, Johnson, and Podgorney}}]{Elphic2008}
Elphic, R.~C., P.~Chu, S.~Hahn, M.~R. James, D.~J. Lawrence, T.~H. Prettyman,
  J.~B. Johnson, and R.~K. Podgorney (2008), Surface and downhole prospecting
  tools for planetary exploration: tests of neutron and gamma ray probes,
  \textit{Astrobiology}, \textit{8}(3), 639--52.

\bibitem[{\textit{Ferr\'e et~al.}(1996)\textit{Ferr\'e, Rudolph, and
  Kachanoski}}]{Ferre1996}
Ferr\'e, P.~A., D.~L. Rudolph, and R.~G. Kachanoski (1996), Spatial averaging
  of water content by time domain reflectometry: Implications for twin rod
  probes with and without dielectric coatings, \textit{Water Resources
  Research}, \textit{32}(2), 271--279, \doi{10.1029/95WR02576}.

\bibitem[{\textit{Ferr\'e et~al.}(1998)\textit{Ferr\'e, Knight, Rudolph, and
  Kachanoski}}]{Ferre1998}
Ferr\'e, P.~A., J.~H. Knight, D.~L. Rudolph, and R.~G. Kachanoski (1998), The
  sample areas of conventional and alternative time domain reflectometry
  probes, \textit{Water Resources Research}, \textit{34}(11), 2971--2979,
  \doi{10.1029/98WR02093}.

\bibitem[{\textit{Franz et~al.}(2013{\natexlab{a}})\textit{Franz, Zreda,
  Rosolem, Hornbuckle, Irvin, Adams, Kolb, Zweck, and
  Shuttleworth}}]{Franz2013eco}
Franz, T., M.~Zreda, R.~Rosolem, B.~K. Hornbuckle, S.~L. Irvin, H.~Adams, T.~E.
  Kolb, C.~Zweck, and W.~J. Shuttleworth (2013{\natexlab{a}}), {Ecosystem-scale
  measurements of biomass water using cosmic ray neutrons}, \textit{Geophysical
  Research Letters}, \textit{40}(1936), \doi{10.1002/grl.50791}.

\bibitem[{\textit{Franz et~al.}(2012{\natexlab{a}})\textit{Franz, Zreda,
  Rosolem, and Ferr\'e}}]{Franz2012}
Franz, T.~E., M.~Zreda, R.~Rosolem, and T.~Ferr\'e (2012{\natexlab{a}}), {Field
  Validation of a Cosmic-Ray Neutron Sensor Using a Distributed Sensor
  Network}, \textit{Vadose Zone Journal}, \textit{11}(4),
  \doi{10.2136/vzj2012.0046}.

\bibitem[{\textit{Franz et~al.}(2012{\natexlab{b}})\textit{Franz, Zreda,
  Ferr\'e, Rosolem, Zweck, Stillman, Zeng, and Shuttleworth}}]{Franz2012a}
Franz, T.~E., M.~Zreda, T.~P.~A. Ferr\'e, R.~Rosolem, C.~Zweck, S.~Stillman,
  X.~Zeng, and W.~J. Shuttleworth (2012{\natexlab{b}}), Measurement depth of
  the cosmic ray soil moisture probe affected by hydrogen from various sources,
  \textit{Water Resources Research}, \textit{48}(8),
  \doi{10.1029/2012WR011871}.

\bibitem[{\textit{Franz et~al.}(2013{\natexlab{b}})\textit{Franz, Zreda,
  Ferr\'e, and Rosolem}}]{Franz2013}
Franz, T.~E., M.~Zreda, T.~P.~A. Ferr\'e, and R.~Rosolem (2013{\natexlab{b}}),
  An assessment of the effect of horizontal soil moisture heterogeneity on the
  area-average measurement of cosmic-ray neutrons, \textit{Water Resources
  Research}, \textit{49}(10), 6450--6458, \doi{10.1002/wrcr.20530}.

\bibitem[{\textit{Franz et~al.}(2013{\natexlab{c}})\textit{Franz, Zreda,
  Rosolem, and Ferr\'e}}]{Franz2013u}
Franz, T.~E., M.~Zreda, R.~Rosolem, and T.~P.~A. Ferr\'e (2013{\natexlab{c}}),
  A universal calibration function for determination of soil moisture with
  cosmic-ray neutrons, \textit{Hydrology and Earth System Sciences},
  \textit{17}(2), 453--460, \doi{10.5194/hess-17-453-2013}.

\bibitem[{\textit{Glasstone and Edlund}(1952)}]{GlasstoneEdlund1952}
Glasstone, S., and M.~C. Edlund (1952), \textit{The elements of nuclear reactor
  theory.}, vii, 416 p. pp., Van Nostrand, New York, includes index.

\bibitem[{\textit{Goldhagen et~al.}(2002)\textit{Goldhagen, Reginatto, Kniss,
  Wilson, Singleterry, Jones, and Steveninck}}]{Goldhagen2002}
Goldhagen, P., M.~Reginatto, T.~Kniss, J.~Wilson, R.~Singleterry, I.~Jones, and
  W.~V. Steveninck (2002), Measurement of the energy spectrum of cosmic-ray
  induced neutrons aboard an {ER-2} high-altitude airplane, \textit{Nuclear
  Instruments and Methods in Physics Research Section A: Accelerators,
  Spectrometers, Detectors and Associated Equipment}, \textit{476}(1–2), 42
  -- 51, \doi{http://dx.doi.org/10.1016/S0168-9002(01)01386-9}.

\bibitem[{\textit{Goldhagen et~al.}(2004)\textit{Goldhagen, Clem, and
  Wilson}}]{Goldhagen02004altlat}
Goldhagen, P., J.~Clem, and J.~Wilson (2004), The energy spectrum of cosmic-ray
  induced neutrons measured on an airplane over a wide range of altitude and
  latitude, \textit{Radiation Protection Dosimetry}, \textit{110}(1-4),
  387--392, \doi{10.1093/rpd/nch216}.

\bibitem[{\textit{Gosse and Phillips}(2001)}]{GossePhillips2001}
Gosse, J.~C., and F.~M. Phillips (2001), Terrestrial in situ cosmogenic
  nuclides: theory and application, \textit{Quaternary Science Reviews},
  \textit{20}(14), 1475 -- 1560,
  \doi{http://dx.doi.org/10.1016/S0277-3791(00)00171-2}.

\bibitem[{\textit{Grimani et~al.}(2011)\textit{Grimani, Araújo, Fabi, Lobo,
  Mateos, Shaul, Sumner, and Wass}}]{Grimani2011}
Grimani, C., H.~M. Araújo, M.~Fabi, A.~Lobo, I.~Mateos, D.~N.~A. Shaul, T.~J.
  Sumner, and P.~Wass (2011), Galactic cosmic-ray energy spectra and expected
  solar events at the time of future space missions, \textit{Classical and
  Quantum Gravity}, \textit{28}(9), 094,005.

\bibitem[{\textit{Gudima et~al.}(1983)\textit{Gudima, Mashnik, and
  Toneev}}]{CEM1983}
Gudima, K., S.~Mashnik, and V.~Toneev (1983), Cascade-exciton model of nuclear
  reactions, \textit{Nuclear Physics A}, \textit{401}(2), 329 -- 361,
  \doi{http://dx.doi.org/10.1016/0375-9474(83)90532-8}.

\bibitem[{\textit{Han et~al.}(2014)\textit{Han, Jin, Li, and Wang}}]{Han2014}
Han, X., R.~Jin, X.~Li, and S.~Wang (2014), Soil moisture estimation using
  cosmic-ray soil moisture sensing at heterogeneous farmland,
  \textit{Geoscience and Remote Sensing Letters, IEEE}, \textit{11}(9),
  1659--1663, \doi{10.1109/LGRS.2014.2314535}.

\bibitem[{\textit{Hands et~al.}(2009)\textit{Hands, Dyer, and Lei}}]{Hands2009}
Hands, A., C.~S. Dyer, and F.~Lei (2009), {SEU Rates in Atmospheric
  Environments: Variations Due to Cross-Section Fits and Environment Models},
  \textit{IEEE Transactions on Nuclear Science}, \textit{56}, 2026--2034,
  \doi{10.1109/TNS.2009.2013466}.

\bibitem[{\textit{Hawdon et~al.}(2014)\textit{Hawdon, McJannet, and
  Wallace}}]{Hawdon2014}
Hawdon, A., D.~McJannet, and J.~Wallace (2014), Calibration and correction
  procedures for cosmic-ray neutron soil moisture probes located across
  {A}ustralia, \textit{Water Resources Research}, \textit{50}(6), 5029--5043,
  \doi{10.1002/2013WR015138}.

\bibitem[{\textit{{Hertel} and {Davidson}}(1985)}]{hertel1985}
{Hertel}, N.~E., and J.~W. {Davidson} (1985), {The response of {B}onner spheres
  to neutrons from thermal energies to 17.3 {MeV}}, \textit{Nuclear Instruments
  and Methods in Physics Research A}, \textit{238}, 509--516,
  \doi{10.1016/0168-9002(85)90494-2}.

\bibitem[{\textit{Huisman et~al.}(2003)\textit{Huisman, Hubbard, Redman, and
  Annan}}]{Huisman2003}
Huisman, J.~A., S.~S. Hubbard, J.~D. Redman, and A.~P. Annan (2003), Measuring
  soil water content with ground penetrating radar: A review, \textit{Vadose
  Zone Journal}, \textit{2}(4), 476--491, \doi{10.2113/2.4.476}.

\bibitem[{\textit{Kazama and Okubo}(2009)}]{KazamaOkubo2009}
Kazama, T., and S.~Okubo (2009), Hydrological modeling of groundwater
  disturbances to observed gravity: Theory and application to asama volcano,
  central japan, \textit{Journal of Geophysical Research: Solid Earth},
  \textit{114}(B8), \doi{10.1029/2009JB006391}.

\bibitem[{\textit{Korres et~al.}(2015)\textit{Korres, Reichenau, Fiener,
  Koyama, Bogena, Cornelissen, Baatz, Herbst, Diekkrüger, Vereecken, and
  Schneider}}]{korres2015}
Korres, W., T.~Reichenau, P.~Fiener, C.~Koyama, H.~Bogena, T.~Cornelissen,
  R.~Baatz, M.~Herbst, B.~Diekkrüger, H.~Vereecken, and K.~Schneider (2015),
  Spatio-temporal soil moisture patterns – a meta-analysis using plot to
  catchment scale data, \textit{Journal of Hydrology}, \textit{520}(0), 326 --
  341, \doi{http://dx.doi.org/10.1016/j.jhydrol.2014.11.042}.

\bibitem[{\textit{Kouzes et~al.}(2008)\textit{Kouzes, Siciliano, Ely, Keller,
  and McConn}}]{Kouzes2007}
Kouzes, R.~T., E.~R. Siciliano, J.~H. Ely, P.~E. Keller, and R.~J. McConn
  (2008), Passive neutron detection for interdiction of nuclear material at
  borders, \textit{Nuclear Instruments and Methods in Physics Research Section
  A: Accelerators, Spectrometers, Detectors and Associated Equipment},
  \textit{584}(2–3), 383 -- 400,
  \doi{http://dx.doi.org/10.1016/j.nima.2007.10.026}.

\bibitem[{\textit{Kowatari et~al.}(2005)\textit{Kowatari, Nagaoka, Satoh, Ohta,
  Abukawa, Tachimori, and Nakamura}}]{Kowatari2005}
Kowatari, M., K.~Nagaoka, S.~Satoh, Y.~Ohta, J.~Abukawa, S.~Tachimori, and
  T.~Nakamura (2005), Evaluation of the altitude variation of the cosmic-ray
  induced environmental neutrons in the mt. fuji area, \textit{Journal of
  Nuclear Science and Technology}, \textit{42}(6), 495--502,
  \doi{10.1080/18811248.2004.9726416}.

\bibitem[{\textit{Kuzhevskij et~al.}(2003)\textit{Kuzhevskij, Nechaev, Sigaeva,
  and Zakharov}}]{Kuzhevski2003}
Kuzhevskij, B., O.~Y. Nechaev, E.~Sigaeva, and V.~Zakharov (2003), Neutron flux
  variations near the earth's crust. a possible tectonic activity detection,
  \textit{Natural Hazards and Earth System Science}, \textit{3}(6), 637--645,
  \doi{10.5194/nhess-3-637-2003}.

\bibitem[{\textit{Larson et~al.}(2008)\textit{Larson, Small, Gutmann, Bilich,
  Braun, and Zavorotny}}]{Larson2008}
Larson, K.~M., E.~E. Small, E.~D. Gutmann, A.~L. Bilich, J.~J. Braun, and V.~U.
  Zavorotny (2008), Use of {GPS} receivers as a soil moisture network for water
  cycle studies, \textit{Geophysical Research Letters}, \textit{35}(24),
  n/a--n/a, \doi{10.1029/2008GL036013}.

\bibitem[{\textit{Lawrence et~al.}(2003)\textit{Lawrence, Elphic, Feldman,
  Prettyman, Gasnault, and Maurice}}]{Lawrence2003}
Lawrence, D.~J., R.~C. Elphic, W.~C. Feldman, T.~H. Prettyman, O.~Gasnault, and
  S.~Maurice (2003), Small-area {T}horium features on the lunar surface,
  \textit{Journal of Geophysical Research: Planets}, \textit{108}(E9),
  \doi{10.1029/2003JE002050}.

\bibitem[{\textit{Legchenko et~al.}(2002)\textit{Legchenko, Baltassat, Beauce,
  and Bernard}}]{Legchenko2002}
Legchenko, A., J.-M. Baltassat, A.~Beauce, and J.~Bernard (2002), Nuclear
  magnetic resonance as a geophysical tool for hydrogeologists, \textit{Journal
  of Applied Geophysics}, \textit{50}(1–2), 21 -- 46,
  \doi{http://dx.doi.org/10.1016/S0926-9851(02)00128-3}.

\bibitem[{\textit{Lei et~al.}(2005)\textit{Lei, Hands, Clucas, Dyer, and
  Truscott}}]{Lei2005}
Lei, F., A.~Hands, S.~Clucas, C.~Dyer, and P.~Truscott (2005), Improvements to
  and validations of the {QinetiQ} atmospheric radiation model (qarm), in
  \textit{Radiation and Its Effects on Components and Systems, 2005. RADECS
  2005. 8th European Conference on}, pp. D3--1--D3--8,
  \doi{10.1109/RADECS.2005.4365581}.

\bibitem[{\textit{Letaw and Normand}(1991)}]{Letaw1991}
Letaw, J.~R., and E.~Normand (1991), Guidelines for predicting single-event
  upsets in neutron environments, \textit{Nuclear Science, IEEE Transactions
  on}, \textit{38}(6), 1500--1506, \doi{10.1109/23.124138}.

\bibitem[{\textit{Lifton et~al.}(2014)\textit{Lifton, Sato, and
  Dunai}}]{Lifton2014}
Lifton, N., T.~Sato, and T.~J. Dunai (2014), Scaling in situ cosmogenic nuclide
  production rates using analytical approximations to atmospheric cosmic-ray
  fluxes, \textit{Earth and Planetary Science Letters}, \textit{386}(0), 149 --
  160, \doi{http://dx.doi.org/10.1016/j.epsl.2013.10.052}.

\bibitem[{\textit{Lin et~al.}(2012)\textit{Lin, Jr., Barghouty, Randeniya,
  Tripathi, Watts, and Yepes}}]{Lin2012}
Lin, Z., J.~A. Jr., A.~Barghouty, S.~Randeniya, R.~Tripathi, J.~Watts, and
  P.~Yepes (2012), Comparisons of several transport models in their predictions
  in typical space radiation environments, \textit{Advances in Space Research},
  \textit{49}(4), 797 -- 806,
  \doi{http://dx.doi.org/10.1016/j.asr.2011.11.025}.

\bibitem[{\textit{Lubczynski and Roy}(2004)}]{LubczynskiRoy2004}
Lubczynski, M., and J.~Roy (2004), Magnetic resonance sounding: New method for
  ground water assessment, \textit{Ground Water}, \textit{42}(2), 291--309,
  \doi{10.1111/j.1745-6584.2004.tb02675.x}.

\bibitem[{\textit{Lv et~al.}(2014)\textit{Lv, Franz, Robinson, and
  Jones}}]{Lv2014}
Lv, L., T.~E. Franz, D.~A. Robinson, and S.~B. Jones (2014), Measured and
  modeled soil moisture compared with cosmic-ray neutron probe estimates in a
  mixed forest, \textit{Vadose Zone Journal}, \textit{13}(12), --.

\bibitem[{\textit{Maurice et~al.}(2004)\textit{Maurice, Lawrence, Feldman,
  Elphic, and Gasnault}}]{Maurice2004}
Maurice, S., D.~J. Lawrence, W.~C. Feldman, R.~C. Elphic, and O.~Gasnault
  (2004), Reduction of neutron data from lunar prospector, \textit{Journal of
  Geophysical Research: Planets}, \textit{109}(E7), \doi{10.1029/2003JE002208}.

\bibitem[{\textit{McJannet et~al.}(2014)\textit{McJannet, Franz, Hawdon,
  Boadle, Baker, Almeida, Silberstein, Lambert, and Desilets}}]{McJannet2014}
McJannet, D., T.~Franz, A.~Hawdon, D.~Boadle, B.~Baker, A.~Almeida,
  R.~Silberstein, T.~Lambert, and D.~Desilets (2014), Field testing of the
  universal calibration function for determination of soil moisture with
  cosmic-ray neutrons, \textit{Water Resources Research}, \textit{50}(6),
  5235--5248, \doi{10.1002/2014WR015513}.

\bibitem[{\textit{McKinney et~al.}(2006)\textit{McKinney, Lawrence, Prettyman,
  Elphic, Feldman, and Hagerty}}]{McKinney2006}
McKinney, G.~W., D.~J. Lawrence, T.~H. Prettyman, R.~C. Elphic, W.~C. Feldman,
  and J.~J. Hagerty (2006), {MCNPX} benchmark for cosmic ray interactions with
  the {M}oon, \textit{Journal of Geophysical Research: Planets},
  \textit{111}(E6), \doi{10.1029/2005JE002551}.

\bibitem[{\textit{{Nesterenok}}(2013)}]{Nesterenok2013}
{Nesterenok}, A. (2013), {Numerical calculations of cosmic ray cascade in the
  Earth's atmosphere - Results for nucleon spectra}, \textit{Nuclear
  Instruments and Methods in Physics Research B}, \textit{295}, 99--106,
  \doi{10.1016/j.nimb.2012.11.005}.

\bibitem[{\textit{Palmiotti et~al.}(2007)\textit{Palmiotti, Salvatores,
  Aliberti, {Nuclear Engineering Division}, and {CEA
  Cadarache}}}]{salvatores2007}
Palmiotti, G., M.~Salvatores, G.~Aliberti, {Nuclear Engineering Division}, and
  {CEA Cadarache} (2007), Validation of simulation codes for future systems :
  motivations, approach, and the role of nuclear data, in \textit{Proceedings
  of the Fourth Workshop on Neutron Measurements, Evaluations and
  Applications}, INL.

\bibitem[{\textit{Pfotzer}(1936)}]{Pfotzer1936}
Pfotzer, G. (1936), Dreifachkoinzidenzen der {U}ltrastrahlung aus vertikaler
  {R}ichtung in der {S}tratosph\"are, \textit{Zeitschrift für {P}hysik},
  \textit{102}(1-2), 41--58, \doi{10.1007/BF01336830}.

\bibitem[{\textit{Rinard}(1991)}]{Rinard1991}
Rinard, P. (1991), Neutron interactions with matter, \textit{Passive
  Non-destructive Assay of Nuclear Materials.}, pp. 357--377.

\bibitem[{\textit{Romano and Forget}(2013)}]{openmcRef}
Romano, P., and B.~Forget (2013), The {OpenMC Monte Carlo} particle transport
  code, \textit{Annals of Nuclear Energy}, \textit{51}(0), 274 -- 281,
  \doi{http://dx.doi.org/10.1016/j.anucene.2012.06.040}.

\bibitem[{\textit{Rosolem et~al.}(2013)\textit{Rosolem, Shuttleworth, Zreda,
  Franz, Zeng, and Kurc}}]{Rosolem2013}
Rosolem, R., W.~J. Shuttleworth, M.~Zreda, T.~E. Franz, X.~Zeng, and S.~a. Kurc
  (2013), {The Effect of Atmospheric Water Vapor on Neutron Count in the
  Cosmic-Ray Soil Moisture Observing System}, \textit{Journal of
  Hydrometeorology}, \textit{14}(5), 1659--1671, \doi{10.1175/JHM-D-12-0120.1}.

\bibitem[{\textit{R{\"u}hm et~al.}(2014)\textit{R{\"u}hm, Mares, Pioch,
  Agosteo, Endo, Ferrarini, Rakhno, Rollet, Satoh, and Vincke}}]{Ruehm2014}
R{\"u}hm, W., V.~Mares, C.~Pioch, S.~Agosteo, A.~Endo, M.~Ferrarini, I.~Rakhno,
  S.~Rollet, D.~Satoh, and H.~Vincke (2014), Comparison of {B}onner sphere
  responses calculated by different {Monte Carlo} codes at energies between 1
  {MeV} and 1 {GeV} - potential impact on neutron dosimetry at energies higher
  than 20 {MeV}, \textit{Radiation Measurements}, \textit{67}(0), 24 -- 34,
  \doi{http://dx.doi.org/10.1016/j.radmeas.2014.05.006}.

\bibitem[{\textit{Salvatores et~al.}(1994)\textit{Salvatores, {Nuclear Energy
  Agency}, and {Organisation for Economic Co-Operation and
  Development}}}]{salvatores1994}
Salvatores, M., {Nuclear Energy Agency}, and {Organisation for Economic
  Co-Operation and Development} (1994), A first approach to data needs and
  target accuracies for hybrid systems, in \textit{Intermediate Energy Nuclear
  Data: Models and Codes}, vol.~27, pp. 313--324.

\bibitem[{\textit{Sato and Niita}(2006)}]{sato2006}
Sato, T., and K.~Niita (2006), Analytical functions to predict cosmic-ray
  neutron spectra in the atmosphere, \textit{Radiation Research},
  \textit{166}(3), 544--555, \doi{10.1667/RR0610.1}.

\bibitem[{\textit{Sato et~al.}(2008)\textit{Sato, Yasuda, Niita, Endo, and
  Sihver}}]{Sato2008}
Sato, T., H.~Yasuda, K.~Niita, A.~Endo, and L.~Sihver (2008), Development of
  parma: Phits-based analytical radiation model in the atmosphere,
  \textit{Radiation Research}, \textit{170}(2), 244--259.

\bibitem[{\textit{Sheu and Jiang}(2003)}]{sheu2003}
Sheu, R., and S.~Jiang (2003), Cosmic-ray-induced neutron spectra and effective
  dose rates near air/ground and air/water interfaces in {T}aiwan,
  \textit{Health Physics}, \textit{84}(1), 92--99,
  \doi{10.1097/00004032-200301000-00008}.

\bibitem[{\textit{Shibata et~al.}(2011)\textit{Shibata, Iwamoto, Nakagawa,
  Iwamoto, Ichihara, Kunieda, Chiba, Furutaka, Otuka, Ohsawa, Murata,
  Matsunobu, Zukeran, Kamada, and Katakura}}]{jendlRef}
Shibata, K., O.~Iwamoto, T.~Nakagawa, N.~Iwamoto, A.~Ichihara, S.~Kunieda,
  S.~Chiba, K.~Furutaka, N.~Otuka, T.~Ohsawa, T.~Murata, H.~Matsunobu,
  A.~Zukeran, S.~Kamada, and J.~Katakura (2011), {JENDL}-4.0: A new library for
  nuclear science and engineering, \textit{Journal of Nuclear Science and
  Technology}, \textit{48}(1), 1--30, \doi{10.1080/18811248.2011.9711675}.

\bibitem[{\textit{Shuttleworth et~al.}(2013)\textit{Shuttleworth, Rosolem,
  Zreda, and Franz}}]{Shuttleworth2013}
Shuttleworth, J., R.~Rosolem, M.~Zreda, and T.~Franz (2013), The cosmic-ray
  soil moisture interaction code ({COSMIC}) for use in data assimilation,
  \textit{Hydrology and Earth System Sciences}, \textit{17}(8), 3205--3217,
  \doi{10.5194/hess-17-3205-2013}.

\bibitem[{\textit{Tate et~al.}(2013)\textit{Tate, Moersch, Jun, Hardgrove,
  Mischna, Litvak, Varenikov, Mitrofanov, Behar, Boynton, Deflores, Fedosov,
  Golovin, Harshman, Kozyrev, Malakhov, Milliken, Mokrousov, Nikiforov, Sanin,
  Vostrukhin, and {MSL Science Team}}}]{Tate2013}
Tate, C., J.~Moersch, I.~Jun, C.~Hardgrove, M.~Mischna, M.~Litvak,
  A.~Varenikov, I.~Mitrofanov, A.~Behar, W.~Boynton, L.~Deflores, F.~Fedosov,
  D.~Golovin, K.~Harshman, A.~Kozyrev, A.~Malakhov, R.~Milliken, M.~Mokrousov,
  S.~Nikiforov, A.~Sanin, A.~Vostrukhin, and {MSL Science Team} (2013), Diurnal
  variations in {MSL DAN} passive measurements with atmospheric pressure and
  soil temperature, \textit{44th Lunar and Planetary Science Conference}, p.
  1601.

\bibitem[{\textit{Vereecken et~al.}(2007)\textit{Vereecken, Kasteel,
  Vanderborght, and Harter}}]{Vereecken2007}
Vereecken, H., R.~Kasteel, J.~Vanderborght, and J.~Harter (2007), {U}pscaling
  {H}ydraulic {P}roperties and {S}oil {W}ater {F}low {P}rocesses in
  {H}eterogeneous {S}oils: {A} {R}eview, \textit{Vadose zone journal},
  \textit{6}, 1 -- 28, \doi{10.2136/vzj2006.0055}, record converted from VDB:
  12.11.2012.

\bibitem[{\textit{Vereecken et~al.}(2008)\textit{Vereecken, Huisman, Bogena,
  Vanderborght, Vrugt, and Hopmans}}]{Vereecken2008}
Vereecken, H., J.~A. Huisman, H.~Bogena, J.~Vanderborght, J.~A. Vrugt, and
  J.~W. Hopmans (2008), On the value of soil moisture measurements in vadose
  zone hydrology: A review, \textit{Water Resources Research}, \textit{44}(4),
  n/a--n/a, \doi{10.1029/2008WR006829}.

\bibitem[{\textit{Wagner et~al.}(2007)\textit{Wagner, Bl{\"o}schl, Pampaloni,
  Calvet, Bizzarri, Wigneron, and Kerr}}]{Wagner2007}
Wagner, W., G.~Bl{\"o}schl, P.~Pampaloni, J.-C. Calvet, B.~Bizzarri, J.-P.
  Wigneron, and Y.~Kerr (2007), Operational readiness of microwave remote
  sensing of soil moisture for hydrologic applications, \textit{Nordic
  Hydrology}, \textit{38}(1), 1--20.

\bibitem[{\textit{Western et~al.}(2004)\textit{Western, Zhou, Grayson, McMahon,
  Blöschl, and Wilson}}]{Western2004}
Western, A.~W., S.-L. Zhou, R.~B. Grayson, T.~A. McMahon, G.~Blöschl, and
  D.~J. Wilson (2004), Spatial correlation of soil moisture in small catchments
  and its relationship to dominant spatial hydrological processes,
  \textit{Journal of Hydrology}, \textit{286}(1–4), 113 -- 134,
  \doi{http://dx.doi.org/10.1016/j.jhydrol.2003.09.014}.

\bibitem[{\textit{Zhu et~al.}(2015)\textit{Zhu, Tan, Gao, and Jiao}}]{Zhu2015}
Zhu, Z., L.~Tan, S.~Gao, and Q.~Jiao (2015), Observation on soil moisture of
  irrigation cropland by cosmic-ray probe, \textit{Geoscience and Remote
  Sensing Letters, IEEE}, \textit{12}(3), 472--476,
  \doi{10.1109/LGRS.2014.2346784}.

\bibitem[{\textit{Zreda et~al.}(2008)\textit{Zreda, Desilets, Ferr\'e, and
  Scott}}]{Zreda2008}
Zreda, M., D.~Desilets, T.~P.~A. Ferr\'e, and R.~L. Scott (2008), Measuring
  soil moisture content non-invasively at intermediate spatial scale using
  cosmic-ray neutrons, \textit{Geophysical Research Letters}, \textit{35}(21),
  \doi{10.1029/2008GL035655}.

\bibitem[{\textit{Zreda et~al.}(2012)\textit{Zreda, Shuttleworth, Zeng, Zweck,
  Desilets, Franz, and Rosolem}}]{Zreda2012}
Zreda, M., W.~J. Shuttleworth, X.~Zeng, C.~Zweck, D.~Desilets, T.~Franz, and
  R.~Rosolem (2012), {COSMOS}: The {CO}smic-ray {S}oil {M}oisture {O}bserving
  {S}ystem, \textit{Hydrology and Earth System Sciences}, \textit{16}(11),
  4079--4099, \doi{10.5194/hess-16-4079-2012}.

\bibitem[{\textit{Zweck et~al.}(2013)\textit{Zweck, Zreda, and
  Desilets}}]{Zweck2013}
Zweck, C., M.~Zreda, and D.~Desilets (2013), Snow shielding factors for
  cosmogenic nuclide dating inferred from {Monte Carlo} neutron transport
  simulations, \textit{Earth and Planetary Science Letters}, \textit{379}(0),
  64 -- 71, \doi{http://dx.doi.org/10.1016/j.epsl.2013.07.023}.

\end{thebibliography}
\end{document}